\documentclass[journal=jctcce,manuscript=article,layout=traditional,]{achemso}

\pdftrailerid{}
\usepackage{chemformula} 
\usepackage[T1]{fontenc} 
\usepackage{amsfonts}
\usepackage{amsmath}
\usepackage{bm}
\usepackage{appendix}
\usepackage{mathrsfs}
\usepackage{mathtools}
\usepackage{nomencl}
\usepackage{siunitx}
\usepackage{xspace}
\usepackage{booktabs}
\usepackage{graphicx}
\usepackage{upgreek}
\usepackage{threeparttable}
\usepackage{algorithm}
\usepackage{algpseudocodex}
\usepackage{subcaption}
\usepackage{url}
\usepackage{adjustbox}

\SectionNumbersOn

\newcommand{\hs}{\ch{H2S}\xspace}

\DeclareSIUnit\angstrom{\text{Å}}  

\newcommand{\orig}{O53\xspace}
\newcommand{\most}[1]{M#1\xspace}
\newcommand{\least}[1]{L#1\xspace}

\newcommand{\zplane}{($110$)\xspace}
\newcommand{\wplane}{($11\bar{2}0$)\xspace}

\title{Sensitivity analysis for {ReaxFF} reparameterization using the {Hilbert--Schmidt} independence criterion}

\author{Michael {Freitas Gustavo}}
\affiliation[Ghent University]{Center for Molecular Modeling, Ghent University, Ghent, Belgium}
\alsoaffiliation{Software for Chemistry and Materials BV, Amsterdam, the Netherlands}

\author{Matti Hellstr\"om}
\affiliation{Software for Chemistry and Materials BV, Amsterdam, the Netherlands}

\author{Toon Verstraelen}
\email{toon.verstraelen@ugent.be}
\affiliation[Ghent University]{Center for Molecular Modeling, Ghent University, Ghent, Belgium}


\makenomenclature

\begin{document}

\begin{abstract}
    We apply a global sensitivity method, the Hilbert--Schmidt independence criterion (HSIC), to the reparameterization of a Zn/S/H ReaxFF force field to identify the most appropriate parameters for reparameterization.
    Parameter selection remains a challenge in this context as high dimensional optimizations are prone to overfitting and take a long time, but selecting too few parameters leads to poor quality force fields.
    We show that the HSIC correctly and quickly identifies the most sensitive parameters, and that optimizations done using a small number of sensitive parameters outperform those done using a higher dimensional reasonable-user parameter selection.
    Optimizations using only sensitive parameters: 1) converge faster, 2) have loss values comparable to those found with the naive selection, 3) have similar accuracy in validation tests, and 4) do not suffer from problems of overfitting.
    We demonstrate that an HSIC global sensitivity is a cheap optimization pre-processing step that has both qualitative and quantitative benefits which can substantially simplify and speedup ReaxFF reparameterizations.
\end{abstract}

\begin{tocentry}
    \includegraphics{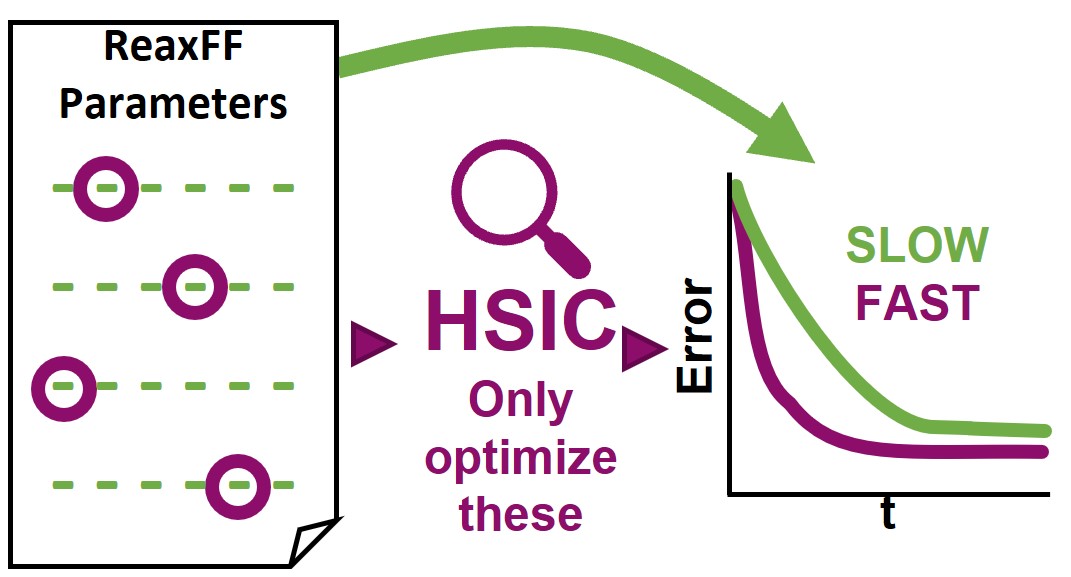}
\end{tocentry}

\newpage
\pagenumbering{arabic}
\section{Introduction}
\label{sec:introduction}

\subsection{ReaxFF reparameterization}
\label{subsec:reaxff-reparameterization}

The ReaxFF (reactive force field)~\cite{VanDuin2001, Senftle2016} is a potential energy surface~(PES) for modelling reactive chemistry at spatial and temporal scales typically unreachable with other, more accurate, but expensive methods.
ReaxFF's speed comes at the cost of replacing accurate formalism with empirical equations that can typically contain hundreds of fitted parameters.
Many of these parameters have no easy physical interpretability and only expert users may have a notion of appropriate values.

This makes the task of fixing the parameters quite difficult.
The procedure typically used to do this is the minimization of some cost function which measures the deviations between predictions made by ReaxFF and some training set of values the user would like to replicate~\cite{Komissarov2021, Gustavo2022, Chenoweth2008, Barcaro2017, Bae2013, Liu2020, Hubin2016, Hu2017, Labrosse2010, Larsson2013}.
However, the question of which parameters should be optimized has always been a difficult one.
Optimizing all potentially relevant parameters simultaneously is prone to producing overfitted results, and very high dimensional optimizations are costly and insufficiently exploratory~\cite{Gustavo2022}.
A key conclusion from our previous work~\cite{Gustavo2022,Shchygol2019} is that the careful conditioning of the error function (in terms of parameter selection and training set items) is crucial to obtaining reasonable results.

Typically, researchers apply rudimentary sensitivity analyses and expert knowledge to guide parameter selection.
However, these processes are usually inaccurate, laborious, or do not sample the space sufficiently.
In this work, we aim to address this problem by introducing a systematic global sensitivity analysis technique to guide parameter selection in a more robust manner.
This method has also been coded into our ParAMS package~\cite{params}, making it a seamless part of any reparameterization workflow.

\subsection{Global sensitivity methods}
\label{subsec:global-sensitivity-methods}

\textit{Uncertainty} analysis concerns itself with the propagation of uncertainty from a model's inputs to its outputs, i.e., estimating an outputs' distribution given all possible changes in the inputs~\cite{Saltelli2019}.
\textit{Sensitivity} analysis attributes the uncertainty of the outputs to particular inputs.
This is broadly done by measuring how much the unconditioned output distribution, $\mathbb{P}(Y)$, differs from the output distribution given certain inputs, $\mathbb{P}(Y|X_i)$~\cite{Baroni2020, Spagnol2019}.
The type of statistical operator used to make this comparison divides global sensitivity methods into two types;
variance-based and distribution-based.

Variance-based methods assume that the variances of these distributions are sufficient to describe them~\cite{Baroni2020,Borgonovo2007}.
This approach is well-established, and the first we tried.
One of the most popular variance-based techniques is known as the Sobol method.
It is based on the assumption that the total variance on the outcomes of a model $Y$ can be linearly apportioned to every input and combination of inputs~\cite{Saltelli2008}:
\begin{equation}
    \label{eq:sobol}
    \sum_{i} S_i + \sum_{i}\sum_{j>i} S_{ij} + \sum_{i}\sum_{j>i}\sum_{k>j} S_{ijk} + \dots + S_{123\dots d} = 1.
\end{equation}
$S_i$ is the Sobol index for the first order effect of parameter $i$ calculated as~\cite{Saltelli2008}:
\begin{equation}
    \label{eq:si}
    S_i = \frac{\mathbb{V}(\mathbb{E}(Y|X_i))}{\mathbb{V}(Y)},
\end{equation}
where $\mathbb{E}$ and $\mathbb{V}$ indicate the expected value and variance respectively.
Higher-order effects are calculated similarly:
\begin{equation}
    \label{eq:sij}
    S_{ij} = \frac{\mathbb{V}(\mathbb{E}(Y|X_i,X_j))}{\mathbb{V}(Y)}.
\end{equation}
A special case is known as the total effect, and is calculated as:
\begin{equation}
    \label{eq:stot}
    S_{Ti} = 1 - \frac{\mathbb{V}(\mathbb{E}(Y|\mathbf{X}_{-i}))}{\mathbb{V}(Y)},
\end{equation}
where $\mathbf{X}_{-i}$ denotes the vector of all parameters except $X_i$.
A full decomposition of the variance is prohibitive for all but the smallest number of dimensions since a total of $2^d - 1$ indices need to be calculated.
It has, however, been shown that the first effect and total effect are sufficient for identifying the most sensitive parameters~\cite{Saltelli2008}.

The main disadvantage of this method is its cost in terms of the number of samples needed.
An efficient method to calculate the first and total effect Sobol indices requires $\eta(d + 2)$ function evaluations~\cite{Saltelli2008}.
There is no prescribed way to select $\eta$, but it is typically on the order of several thousand~\cite{Herman2013,Nossent2011}.
Since the cost of the calculation of the indices themselves is trivial in comparison to the cost of obtaining the samples, the recommended procedure in literature is to iteratively sample and calculate until the indices converge.
In this work, we attempt to introduce a preprocessing step to a ReaxFF parameterization, and thus the cost of this should be significantly less than the optimization;
literature sources suggest that this will not be the case~\cite{Nossent2011,Herman2013}.

The intractable expense of the Sobol method is well-known, and much effort has gone into decreasing its cost~\cite{Saltelli2008}.
One popular method which is significantly cheaper than the Sobol one, is the elementary effects method, also known as Morris screening~\cite{Campolongo2011}.
Morris screening uses a careful sampling strategy to calculate an `estimated effect' which has been shown to be a good proxy for the total effect Sobol index~\cite{Saltelli2008,Campolongo2011}.
This method is a popular screening method for dimensionality reduction in literature, particularly for expensive models.
In theory, Morris screening should be a suitable method for our purposes, however, in our early investigations, this proved to not be the case.

Both Sobol indices and Morris screening impose rules on how samples may be generated.
In the case of Sobol indices, certain parameter values must be fixed while allowing others to vary.
In the case of Morris screening, samples are gathered in trajectories where individual parameters are changed one at a time.
In our particular context, these rules make obtaining an adequate number of valid samples very frustrating and time-consuming.
This is because the ReaxFF loss function surface is full of undefined points corresponding to parameter values which result in crashed calculations and nonphysical results.
If such points are encountered when generating a Morris trajectory, for example, then the entire trajectory must be abandoned because these method do not allow the entry of nonfinite or undefined values.
It would be simpler to find a method which allowed these values, or did not have sampling rules so that such instances could simply be discarded.

Aside from the real, practical issues related to their use, the assumptions made by variance-based techniques are inappropriate in the context of identifying parameter sensitivities in the ReaxFF error function.
This is because it can produce extremely large values, and minima which are often located in narrow valleys and are unlikely to be captured during a sampling procedure~\cite{Gustavo2022}.
Cheaper approximation methods are typically insufficiently space-filling, or rely on strong assumptions about the structure of the inputs or outputs~\cite{DeLozzo2016, Saltelli2019}.
The main criticism of variance-based techniques, however, is the fact that they limit the amount of information extracted from distributions to a single metric.

Distribution-based sensitivity approaches represent a newer family of methods.
Unlike variance-based methods, these techniques do not reduce distributions to only their variance and can, in principle, capture arbitrary dependencies by considering the entirety of the distributions.
Different statistics are used to measure the differences between distributions~\cite{Baroni2020}, and various methods of this type have been developed~\cite{Borgonovo2007, DaVeiga2015, DeLozzo2016}.
We are interested in the Hilbert--Schmidt Independence Criterion~(HSIC)~\cite{Gretton2005} for it simplicity, speed and accuracy.

The HSIC was first introduced in \citet{Gretton2005}.
Through the use of kernels, it can identify arbitrary non-linear dependencies between the inputs and outputs~\cite{Baroni2020}.
It is simpler and faster to converge than many other distribution-based techniques, and does not require explicit regularization~\cite{Gretton2005}.
It has since been used in: feature selection~\cite{Yamada2018, Yamada2014, DaVeiga2015, Song2012, Climente-Gonzalez2019, DeLozzo2016}, event detection~\cite{Feng2018, Qian2022}, machine learning~\cite{Li2021b, Wang2018}, and dimensionality reduction in the context of global optimization~\cite{Spagnol2019}, among other applications.
This final use-case is of particular interest in our context and, other than \citet{Spagnol2019}, we have found no other work that directly applied the HSIC as a pre-optimization step.

In this work, we apply an HSIC analysis to a ReaxFF reparameterization problem.
Based on the sensitivity results, we reduce the original high-dimensional optimization to a lower dimensional one.
We then demonstrate that the optimization in this reduced space is able to produce ReaxFF force fields which are competitive with those produced by the higher dimensional optimization, but converge more quickly and reduce the risk of overfitting.

We focus on ReaxFF because parameterizations of this force field occur regularly in literature~\cite{Kaymak2022,Pahari2012,Hu2017,Trnka2018,Jaramillo2014}.
However, the techniques we introduce here could conceivably be applied to any high-dimensional parameterization problem of the type that occurs frequently in many fields~\cite{Boothroyd2022,Jonsson2022,Mondal2020,Verstraelen2011,He2022,Wang2022,Zhang2020,Gopalakrishnan2020,Mohammed2018}, because they all share the same underlying sloppy mathematical behavior~\cite{Quinn2019}.

In the next section we will introduce ReaxFF, the HSIC metric, the kernels used, and the reparameterization loss function.
In Section~\ref{sec:the-reparameterization-problem} we introduce a test case on which we apply the sensitivity technique, the results of which are described in Section~\ref{sec:results-and-discussion}, and future research paths are discussed in Section~\ref{sec:outlook}.
Section~\ref{sec:software} details our implementation of the calculation, and conclusions are drawn in Section~\ref{sec:conclusions}.

\section{Mathematical derivations}
\label{sec:mathematical-derivations}

\subsection{ReaxFF energy potential}
\label{subsec:reaxff-energy-potential}

The elucidation of the full ReaxFF energy potential is beyond the scope of this work, however, we provide a brief introduction here for context.
Interested readers can consult \citet{Chenoweth2008} and \citet{Senftle2016} for more details.

The ReaxFF energy potential is a summation of different energy contributions~\cite{Chenoweth2008}:
\begin{align}
    E_\textrm{sys} = &E_\textrm{bond} + E_\textrm{lp} + E_\textrm{over} + E_\textrm{under} + E_\textrm{val} \nonumber\\
                     &+ E_\textrm{pen} + E_\textrm{coa} + E_\textrm{C2} + E_\textrm{triple} \nonumber\\
                     &+ E_\textrm{tors} + E_\textrm{conj} + E_\textrm{H-bond} + E_\textrm{VdW} \nonumber\\
                     &+ E_\textrm{charges} + \dots \label{eq:reaxff}
\end{align}

The main contribution is the bonding energy, $E_\textrm{bond}$, which calculates an energy based on bond orders, which are, in turn, calculated from atomic positions (the most fundamental inputs to any PES).
The other contributions can be thought of as corrections to the bonding energy.
Some corrections are general, for example $E_\textrm{over}$ and $E_\textrm{under}$ which penalize atoms which are over- or undercoordinated by their bonding.
Some corrections are extremely specific, for example $E_\textrm{C2}$ which corrects the energy for the C$_2$ molecule.
There is also an energy term for fluctuating charges ($E_\textrm{charges}$).
ReaxFF typically uses either the EEM~\cite{Mortier1986} or ACKS2~\cite{Verstraelen2013} charge models.
In this work we use EEM charges.

Each of the energy contributions is typically a complicated function of bond orders and empirical parameters.
For example~\cite{Chenoweth2008}:
\begin{align}
    E_\textrm{bond} &= -D^{\sigma}_e \cdot BO^{\sigma}_{ij} \cdot \exp
        \left[
            p_{be1}
            \left(
                1 - \left( BO^\sigma_{ij} \right)^{p_{be2}}
            \right)
        \right] \nonumber\\
        &-D^{\pi}_e \cdot BO^{\pi}_{ij}
         -D^{\pi\pi}_e \cdot BO^{\pi\pi}_{ij} \label{eq:bond_energy}
\end{align}
where $D^{\sigma}_e$, $D^{\pi}_e$, $D^{\pi\pi}_e$, $p_{be1}$ and $p_{be2}$ are parameters and $BO_{ij}$ are bond orders.

The complexity and number of contributions quickly leads the ReaxFF PES to have a very large number of parameters, particularly because most parameters are a function of one or more atom types.
For example, there is a $D^\sigma_e$ parameter for every combination of atom types being modelled.

In an effort to make the parameters easier to understand, the original authors organised them into the following groups or blocks: general, atoms, bonds, off-diagonal, angles, torsions and hydrogen bonds.
These groups have no bearing on the calculation of the PES, but serve as a helpful organisational tool, and appear in the formatting rules for the inputs of most ReaxFF implementations.
We will return to these groups when we introduce a new Zn/S/H force field in Section~\ref{subsec:initial-force-field}.

We use the term `force field' in this work to refer to a full set of ReaxFF parameters and their values.
Force fields are functions of the atom types they include and are sometimes only appropriate for specific conditions.
For example, ReaxFF force fields containing hydrogen and oxygen are generally classed as appropriate for aqueous or combustion chemistry.

\subsection{Loss function}
\label{subsec:loss-function}

The reparameterization of a force field requires the construction of a training set ($\mathbf{b}_{\text{ref}} \in \mathbb{R}^m$), which is a vector of $m$ scalar training point values of chemical properties that one would like ReaxFF to replicate.
The goal of the reparameterization is the minimization of a loss function which quantifies the overall difference between the ReaxFF model and the training data in a single number.
This loss function can take various forms, but the most common is the sum of square errors (SSE).
We define the SSE as:
\begin{equation}
    \label{eq:sse}
    \mathrm{SSE}(\mathbf{x}) \coloneqq
    \sum^m_{i=1} w_i \left(
    \frac{ b_{\text{ref}, i} - b(\mathbf{x}, \mathbf{G}_i)}{\sigma_i}
    \right)^2.
\end{equation}

Each item in the training set ($b_{\text{ref}, i}$) is compared to the result produced by the ReaxFF model ($b(\mathbf{x}, \mathbf{G}_i)$) using a vector of parameters which we allow to be adjusted ($\mathbf{x}$), and a set of system inputs and fixed parameters ($\mathbf{G}_i$).
We will call the parameters we have chosen to optimize \textit{active} parameters, and denote the set of active parameters $\mathscr{D}$.
Items in the training set can be of different types (energies, forces, angles, etc.).
To make them comparable and independent of units, each term in the loss function is divided by an appropriate $\sigma_i$ with the same unit as $b_{\text{ref}, i}$.
The $\sigma_i$ value is often understood as the acceptable error for that item.
Finally, different training values can be weighted differently based on their importance via the weighting vector ($\mathbf{w}$).
For example, it is more important to get energy minima correct than it is to replicate energies for very exaggerated geometries;
one might weight the former higher than the latter in the loss function.
In the case of multi-dimensional data, like forces, the $n \times 3$ matrix is simply flattened into the vector of training data.
In other words, a single element of $\mathbf{b}_{\text{ref}}$ would not be `the forces on molecule A' for example, but rather `the x-component of the forces on atom 1 in molecule A'.
For more details regarding the implementation itself, readers are directed to the ParAMS documentation~\cite{params}.

$\bm{\upsigma}$ and $\mathbf{w}$ perform essentially similar functions since they are both constants which could conceivably be combined.
Using one, or the other, or both is a question of preference for different practitioners.
Traditionally, only $\bm{\upsigma}$ was used, however for some users, weights provide a more intuitive value.
We have chosen to use a fixed vector of values for $\bm{\upsigma}$, and adjust the weights to balance the training set.

\subsection{Hilbert--Schmidt independence criterion}
\label{subsec:Hilbert--Schmidt-independence-criterion}

This section introduces the Hilbert--Schmidt independence criterion~(HSIC) from a practitioner's perspective.
Our presentation is not as general as in other works,~\cite{Gretton2005, Song2012, Spagnol2019}
and we only strive to explain the basic idea to readers interested in parameter selection.
We will assume the space of a single parameter is $\mathcal{X} \subseteq \mathbb{R}$ and the space of loss values is $\mathcal{Y} \subseteq \mathbb{R}$.
Typically, parameters are bounded and loss values are positive but these are not strict requirements for HSIC\@.

Let the Hilbert space of $\mathbb{R}$-valued functions on $\mathcal{X}$ be
$\mathcal{H}_x : \mathcal{X} \rightarrow \mathbb{R}$,
for which an inner product $\langle \cdot, \cdot \rangle_{\mathcal{H}_x}$ is defined.
This is called a \textit{reproducing kernel} Hilbert space~(RKHS) when
a kernel $k : \mathcal{X} \times \mathcal{X} \rightarrow \mathbb{R}$ exists such that $\forall x \in \mathcal{X}$ and $\forall f \in \mathcal{H}_x$:
\begin{align*}
    k(x, \cdot) &\in \mathcal{H}_x\text{ and }\\
    \left\langle f, k(x,\cdot) \right\rangle_{\mathcal{H}_x} &= f(x).
\end{align*}
The second condition is known as the \textit{reproducing} property: a function evaluation $f(x)$ can be reproduced by taking an inner product of $f$ with a partially evaluated kernel.
We similarly define the RKHS, $\mathcal{H}_y$ with kernel $\ell$, for for the loss-value space $\mathcal{Y}$.

The kernel $k$ can be defined in terms of a feature map $\phi : \mathcal{X} \rightarrow \mathcal{F}_x$,
where $\mathcal{F}_x$ is a new Hilbert space, often called the feature space of $x$.
This means that each value $x$ is uniquely represented by a function $\phi(x) \in \mathcal{F}_x$.
Any positive definite and symmetric kernel $k$ can always be described as an inner product in feature space:
\begin{align}
    \label{eq:k2}
    k(x, x') &\coloneqq \left\langle \phi(x), \phi(x') \right\rangle_{\mathcal{F}_x}
\end{align}
In practice, direct use of $\phi$ is avoided
and the kernel $k$ is used instead to keep calculations manageable.
Kernels are generally cheap to evaluate, even when the feature map would be computationally infeasible.
Similarly, a feature map $\psi : \mathcal{Y} \rightarrow \mathcal{F}_y$ exists, such that $\ell(y, y') \coloneqq \left\langle \psi(y), \psi(y') \right\rangle_{\mathcal{F}_y}$.

Now consider random variables $X \in \mathcal{X}$ and $Y \in \mathcal{Y}$,
with a joint distribution $\mathbb{P}_{XY}$.
Variance-based sensitivity methods characterize the dependence of $Y$ on $X$
with $\operatorname{cov}[X,Y]=\mathbb{E}_{XY}[XY]-\mathbb{E}_X[X]\mathbb{E}_Y[Y]$,
which is only picking up linear correlations.
In principle, one may overcome this limitation by computing the covariance in feature space instead:
$\operatorname{cov}[\phi(X),\psi(Y)]=\mathbb{E}_{XY}[\phi(X)\psi(Y)]-\mathbb{E}_X[\phi(X)]\mathbb{E}_Y[\psi(Y)]$.
However, the latter covariance is impractical because it directly operates in feature space
and because the covariance is an element of a new Hilbert space $\mathcal{F}_x\otimes\mathcal{F}_y$,
which troubles its direct interpretation.
\citet{Gretton2005} showed how to overcome both issues by rewriting the squared norm of the latter covariance in terms of kernels:
\begin{align}
    \text{HSIC}(\mathcal{F}_x, \mathcal{F}_y, \mathbb{P}_{xy})
    &\coloneqq \bigl\Vert\operatorname{cov}[\phi(X),\psi(Y)]\bigr\Vert^2_{\mathcal{F}_x\otimes\mathcal{F}_y} \\
    &=\begin{aligned}[t]
      &\mathbb{E}_{XX'YY'}[k(X,X')\ell(Y,Y')]\\
      &+\mathbb{E}_{XX'}[k(X,X')]
        \mathbb{E}_{YY'}[\ell(Y,Y')]\\
      &-2 \mathbb{E}_{XY}\bigl[
          \mathbb{E}_{X'}[k(X,X')]
          \mathbb{E}_{Y'}[\ell(Y,Y')]
        \bigr]
    \end{aligned}
\end{align}
This definition of HSIC yields a positive real number, which increases as $X$ and $Y$ are more correlated.
When $k$ and $\ell$ are linear kernels, HSIC reduces to $\bigl\vert\operatorname{cov}[X,Y]\bigr\vert^2$ and there is no added value compared to variance-based sensitivity criteria.
In case of non-linear kernels, the HSIC also detects non-linear correlations between $X$ and $Y$.
Ideally, $k$ and $\ell$ are \textit{characteristic kernels},~\cite{Spagnol2019} for which $\text{HSIC}=0$ if and only if $X$ and $Y$ are independent ($\mathbb{P}_{XY} = \mathbb{P}_X \mathbb{P}_Y$).
Of the common kernels, the Gaussian kernel is characteristic, while the linear and polynomial kernels are not.
It is possible to use non-characteristic kernels for HSIC, but then one loses the guarantee that all dependence will be captured~\cite{Song2012}.

It is usually not possible to obtain $\mathbb{P}_{xy}$ explicitly, however, the HSIC can be estimated if we sample the distribution.
We define a set of samples as:
\begin{equation}
    \label{eq:def_z}
    Z \coloneqq (\mathbf{X},\mathbf{y}),
\end{equation}
where $\mathbf{X} \in \mathbb{R}^{n \times d}$ is a matrix of $n$ rows of uniformly randomly sampled parameter vectors $\{\mathbf{X}_1, \mathbf{X}_2, \dots \mathbf{X}_n\}$, and we let $\mathbf{y} \in \mathbb{R}^n = \{ \mathrm{SSE}(\mathbf{X}_1), \mathrm{SSE}(\mathbf{X}_2) \dots, \mathrm{SSE}(\mathbf{X}_n) \}$ be the corresponding set of loss function values.

To measure the sensitivity of a parameter  with index $\omega$, we take the associated column with $n$ sample values, $\mathbf{X}_{\cdot\omega}$, and the loss values:
\begin{equation}
    \label{eq:def_Zw}
    Z_\omega \coloneqq (\mathbf{X}_{\cdot\omega},\mathbf{y}),
\end{equation}
and we make use of the \textit{unbiased} estimator of the HSIC from \citet{Song2012}:
\begin{align}
    \label{eq:hsic_est}
    \mathrm{HSIC}(\mathcal{F}_x, \mathcal{F}_y, Z_\omega) \approx
    \frac{1}{n(n-3)}
    \left[
        \rho_1
        + \rho_2
        - \rho_3
    \right],
\end{align}
where:
\begin{align*}
    \rho_1
    &\coloneqq
    \mathrm{tr}(\mathbf{\tilde{K}}\mathbf{\tilde{L}})
    \\
    \rho_2
    &\coloneqq
    \frac{\bm{1}^\top\mathbf{\tilde{K}}\bm{1}\bm{1}^\top\mathbf{\tilde{L}}\bm{1}}{(n-1)(n-2)}
    \\
    \rho_3
    &\coloneqq
    \frac{2}{n-2} \bm{1}^\top\mathbf{\tilde{K}}\mathbf{\tilde{L}}\bm{1}
    \\
    \mathbf{\tilde{K}}_{ij}
    &\coloneqq
    (1 - \delta_{ij})k(x_i, x_j)
    \quad\forall x_i, x_j \in X_{\cdot\omega}
    \\
    \mathbf{\tilde{L}}_{ij}
    &\coloneqq (1 - \delta_{ij})\ell(y_i, y_j)
    \quad\forall y_i, y_j \in \mathbf{y}.
\end{align*}

$\mathrm{HSIC}$ is a convenient metric with which to rank the influence of parameters on an output, however, to make interpretation even easier we define the sensitivity of a parameter as the normalized HSIC values~\cite{Spagnol2019}:
\begin{equation}
    \label{eq:sensitivity}
    s_{\omega} =
    \frac{\mathrm{HSIC}(\mathcal{F}_x, \mathcal{F}_y, Z_\omega)}
    {\sum^d_{i=1} \mathrm{HSIC}(\mathcal{F}_x, \mathcal{F}_y, Z_i)}.
\end{equation}
By construction, sensitivity values are bounded ($0 \leq s_\omega \leq 1$) and $\sum s_\omega = 1$.
This makes sensitivities a more intuitive value to interpret and understand than $\mathrm{HSIC}$.

The advantage of the \textit{biased} HSIC estimator~\cite{Song2012} is that it is positive by construction.
The \textit{unbiased} estimator used here, although more accurate, runs the risk of producing small negative values for low sensitivity items.
In the context of the sensitivity analysis, where a qualitative consideration of the results is arguably more important that a quantitative one, we simply clip any negative HSIC values to zero before calculating the sensitivities (Equation~\ref{eq:sensitivity}).

For a more detailed description of HSIC, the interested reader is directed to \citet{Gretton2005} for the original introduction, and to \citet{DaVeiga2015, Song2012, DeLozzo2016} and \citet{Spagnol2019} for more recent works.

Note that it is also possible to consider a more fine-grained sensitivity approach and let $\mathbf{y}$ be individual residual values (i.e., $y_i = b_{\text{ref}, i} - b(\mathbf{x}, \mathbf{G}_i))$.
This would allow one to determine sensitivities for particular training set items;
further information to assist in training set design.
This represents an extension of the current work, and will not be dealt with here, but is left as an avenue for further study.

\subsection{Kernels}
\label{subsec:kernels}

\begin{figure}
    \centering
    \includegraphics{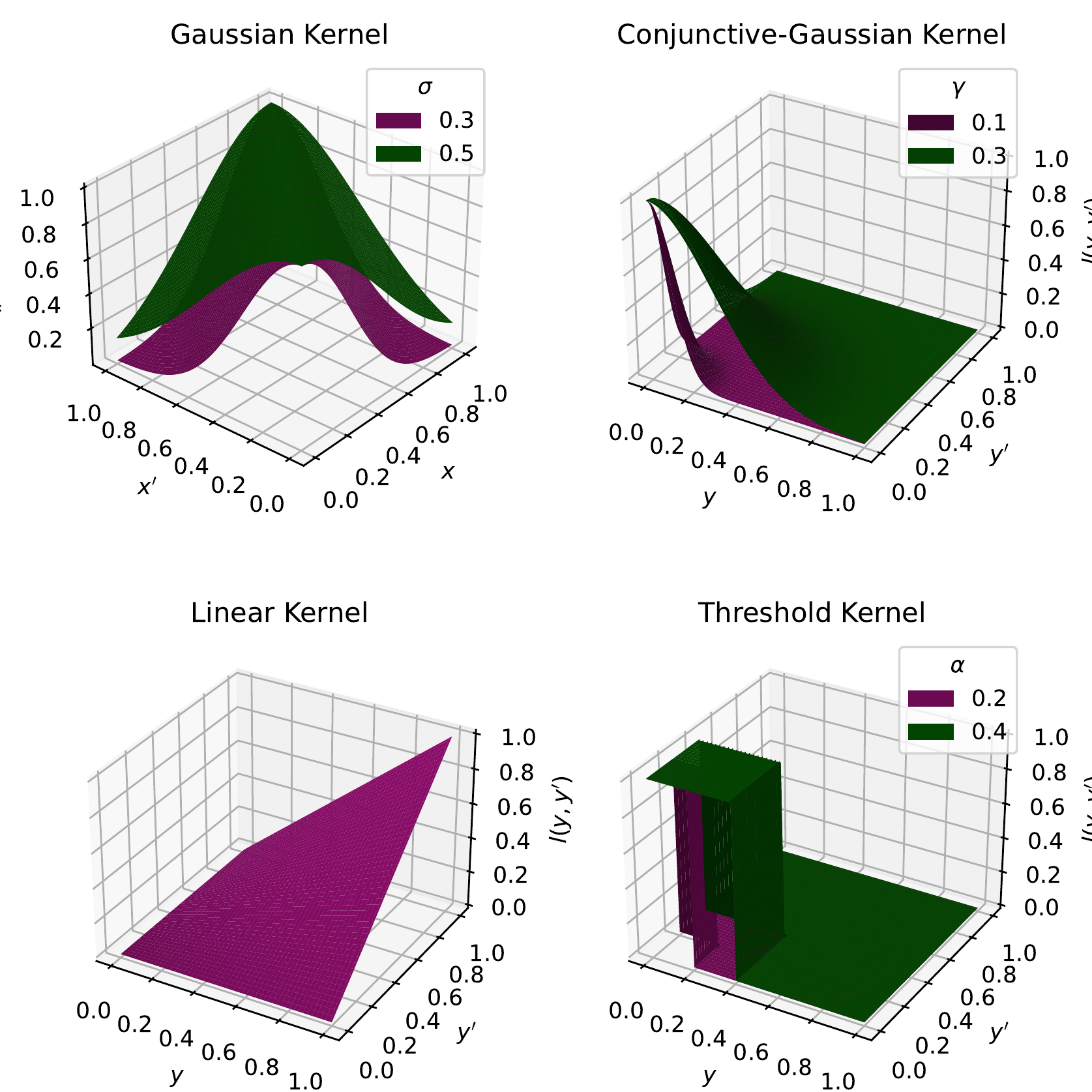}
    \caption{Plots of the tested kernel functions.}
    \label{fig:kernels}
\end{figure}

Figure~\ref{fig:kernels} illustrates the kernel distributions used in this work, using different parameter values where appropriate.

We select the characteristic Gaussian kernel for the parameters-kernel:
\begin{equation}
    \label{eq:gaussian_kernel}
    k(x, x') \coloneqq \exp\left[ -\frac{(x - x')^2}{2\sigma^2} \right].
\end{equation}
This captures the dis/similarity between parameters, and allows the distribution to take an arbitrary shape.
For the sake of clarity, $\sigma$ as it is used in Equation~\ref{eq:gaussian_kernel} is unrelated to the same symbol in Equation~\ref{eq:sse}.

We test several candidates for the loss-kernel.
We first consider the non-characteristic threshold kernel introduced by \citet{Spagnol2019}:
\begin{align}
    \label{eq:threshold}
    \ell(y, y') &\coloneqq h(y) h(y') \\
    h(y) &\coloneqq
    \begin{cases}
        1 &\text{ if } y < q_{\alpha}, \\
        0 &\text{ otherwise}.
    \end{cases}
\end{align}
$q_{\alpha}$ is the value corresponding to the $\alpha$ quantile of the data.
The above kernel is constructed to focus the weight of the distribution on the smallest numerical values of $\mathbf{y}$ since these represent good loss function values we would like the most information about.
\citet{Spagnol2019} showed that thresholding was extremely important in identifying sensitivities near minima, and not allowing the calculation to be swamped by very large function values.

We then consider a characteristic and continuous approximation of the threshold kernel, which we call the conjunctive-Gaussian~(CG) kernel:
\begin{equation}
    \label{eq:cg_kernel}
    \ell(y, y') \coloneqq \exp\left[ -\frac{(y^2 + y'^2)}{2\gamma^2} \right],
\end{equation}
where lower values of the $\gamma$ parameter focus the distribution more heavily on the best values.

For comparison, we also test a simple non-characteristic linear kernel:
\begin{equation}
    \label{eq:linear}
    \ell(y, y') \coloneqq y y',
\end{equation}
and the Gaussian kernel introduced in Equation~\ref{eq:gaussian_kernel}.
Other kernels are possible, but we leave their consideration for later work.

To stabilize the numerics and handle order-of-magnitude concerns, we transform and scale the loss values before applying the loss-kernel to them:

\begin{equation}
    \label{eq:def_ytilde}
    \tilde{y}(y) = \frac{\ln{y} - \min{(\ln{\mathbf{y}})}}{\max{(\ln{\mathbf{y}})} - \min{(\ln{\mathbf{y}})}}.
\end{equation}

\section{Software}
\label{sec:software}

Training data was calculated using the Amsterdam Modeling Suite~(AMS) 2022.

The sensitivity calculation has been implemented in the Python programming language, and integrated into a development version of ParAMS~2023~\cite{params}, a reparameterization toolkit which comes bundled with AMS\@.
It is also used to handle the force field optimizations through the GloMPO optimization management software~\cite{Gustavo2022}.

\section{The reparameterization problem}
\label{sec:the-reparameterization-problem}

Our main goal in this work is to demonstrate the method and utility of our new sensitivity approach.
For this reason, the training set we have designed is smaller than what would typically be required to create a production quality parameterization.
Designing a large training set would require an entire publication in its own right, and distract heavily from the focus of our article.
A smaller training set allows us to keep the discussions more focused and clear.

In any case, training set design is always an iterative process where more items are added as validation tests on the new force fields demonstrate deficiencies in the original set~\cite{Muller2016}.
This work can be considered a first iteration upon which later work can build.

\subsection{Initial force field}
\label{subsec:initial-force-field}

We aim to create a new ReaxFF force field which correctly models the adsorption of \hs on ZnS -- an absorbent which has received some attention for its favorable electronic properties~\cite{Dloczik2001, Hamad2002, Li2021, Raymand2010, Qi2014}.
\hs is a common, but toxic, gas and its adsorption behavior on ZnS has been investigated for both gas detection~\cite{Qi2014} and gas removal purposes~\cite{Li2021}.
As far as we are aware, however, no specially designed Zn/S/H ReaxFF force field exists to model this behaviour.

Table~\ref{tab:blocknames} details the nomenclature we will use to refer to the various parameter blocks which compose the ReaxFF force field file~\cite{ffieldformat}.
For a discussion of the parameter blocks, see Section~\ref{subsec:reaxff-energy-potential}

\begin{table}
    \centering
    \caption{Aliases used to refer to different parameter blocks in the ReaxFF force field file format.}
    \label{tab:blocknames}
    \begin{tabular}{lll}
        \toprule
        Alias & Description & Number of parameters \\
        \midrule
        GEN         & General parameters & 41 \\
        ATM:W       & Parameters for atoms of type W & 32 per atom type \\
        BND:W.X     & Parameters for W-X bonds & 16 per W.X pair \\
        OFD:W.X     & Off-diagonal definitions for atom pairs & 6 per W.X pair \\
        ANG:W.X.Y   & Parameters for angles formed by W-X-Y atoms & 7 per angle group \\
        TOR:W.X.Y.Z & Parameters for torsions formed by W-X-Y-Z atoms & 7 per angle group \\
        HBD:W.H.X   & Hydrogen-bonding parameters for atoms W and X & 4 per W.X pair \\
        \bottomrule
    \end{tabular}
\end{table}

Our initial force field is an amalgamation of parameter values from already published force fields:
\begin{enumerate}
    \item The ATM:S, BND:S.S and ANG:S.S.S parameter blocks are filled with values from the Li/S force field of \citet{Islam2015};
    \item ATM:H, BND:S.H and ANG:H.S.H values are taken from the C/H/O/S force field of \citet{Muller2016};
    \item GEN, ATM:Zn, BND:H.H, BND:Zn.H, BND:Zn.Zn, BND:Zn.S, ANG:Zn.Zn.S, ANG:Zn.S.Zn, ANG:Zn.S.S, ANG:S.Zn.S and HBD:S.H.S blocks are filled with parameters published in the Zn/O/H force field of \citet{Raymand2010}.
    This force field does not contain any sulfur compounds, so the corresponding oxygen-related parameter blocks are used.
    For example, BND:Zn.O parameters are filled into the BND:Zn.S block of the initial force field.
\end{enumerate}

The force field contains \num{279} parameters.
From this we use some common intuition -- based on the contents of the training set and descriptions of the parameters -- to select \num{53} for initial optimization.
The selection is `greedy' in that a parameter is selected for optimization if:
\begin{enumerate}
    \item it could conceivably affect the behaviour of a training set item,
    \item the training set contains items related to the parameter, and
    \item the parameter is appropriate for optimization.
\end{enumerate}
For example, the atomic masses of the elements are technically parameters, however, changing them would not be appropriate.
Similarly, $\pi$-bonds are not present in the training set, thus, their related parameters are not chosen for optimization.
Finally, we do not optimize any atomic or general block parameters.
These are generally known to be more `expert' level parameters that are not suitable for preliminary stage optimization~\cite{params}.

The \num{53} parameters selected for optimization are listed in Table~\ref{tab:initial_selection}.
The default parameter ranges supplied by ParAMS~\cite{params} were used, and can be found in the supplementary information.

\begin{table}
    \centering
    \caption{Initial naive parameter selection. Equation numbers and parameter names refer to notation used in~\cite{scmdevdocs}.}
    \adjustbox{max width=\linewidth}{%
    \begin{tabular}{lllll}
        \toprule
        \textbf{Block} & \textbf{Name} & \textbf{Eqn.} & \textbf{Description} & \textbf{Atoms} \\
        \midrule
        ANG &\texttt{-p\_hb2}    & 18     & Hydrogen bond/bond order           & H.S.H \\
            &\texttt{-p\_hb3}    & 18     & Hydrogen bond parameter            & H.S.H \\
            &\texttt{p\_hb1 }    & 2      & Hydrogen bond energy               & H.S.H \\
            &\texttt{p\_val1}    & 13a    & Valance angle parameter            & H.S.H S.S.Zn S.Zn.S S.Zn.Zn Zn.S.Zn \\
            &\texttt{p\_val2}    & 13a    & Valance angle parameter            & H.S.H S.S.Zn S.Zn.S S.Zn.Zn Zn.S.Zn \\
            &\texttt{p\_val4}    & 13b    & Valance angle parameter            & H.S.H S.S.Zn S.Zn.S S.Zn.Zn Zn.S.Zn \\
            &\texttt{p\_val7}    & 13c    & Under-coordination                 & H.S.H S.S.Zn S.Zn.S S.Zn.Zn Zn.S.Zn \\
            &\texttt{Theta\_0,0} & 13g    & 180\textdegree-(equilibrium angle)    & H.S.H S.S.Zn S.Zn.S S.Zn.Zn Zn.S.Zn \\
        HBD &\texttt{r\_hb\textasciicircum 0} & 18     & Hydrogen bond equilibrium distance & S.H.S \\
        BND &\texttt{D\_e\textasciicircum sigma} & 6, 11a & Sigma-bond dissociation energy     & H.S S.S S.Zn Zn.Zn \\
            &\texttt{p\_be1}     & 6      & Bond energy parameter              & H.S S.S S.Zn Zn.Zn \\
            &\texttt{p\_be2}     & 6      & Bond energy parameter              & H.S S.S S.Zn Zn.Zn \\
            &\texttt{p\_bo1}     & 2      & Sigma bond order                   & H.S S.S S.Zn Zn.Zn \\
            &\texttt{p\_bo2}     & 2      & Sigma bond order                   & H.S S.S S.Zn Zn.Zn \\
            &\texttt{p\_ovun1}   & 11a    & Over-coordination penalty          & H.S S.S S.Zn Zn.Zn \\
        \bottomrule
    \end{tabular}%
    }
    \label{tab:initial_selection}
\end{table}

\subsection{Training data}
\label{subsec:training-data}

The training set, calculated with BAND using PBEsol and DZ or DZP basis set, consists of 16 jobs, from which 471 individual training points are extracted.

AMS BAND calculates charges via the Hirshfeld, Voronoi deformation, Mulliken and CM5 methods.
We have included Hirshfeld charges in our training set.
Although these charges are not without problems~\cite{Verstraelen2016}, we believe they are the best available.
The accuracy of these charges is also not overly important since we do not activate any charge related parameters in our initial parameter selection (see Section~\ref{subsec:initial-force-field}), and the few charges we include in the training set are mainly used as a sanity check on the results.

The training set consists of:

\begin{enumerate}
    \item energy-volume scans, charges and enthalpies of formation for the cubic zincblende, cubic rocksalt and hexagonal wurtzite structures of ZnS;
    \item the optimized geometry, charges, H-S bond scan, and H-S-H angle scan for \hs;
    \item the energy of formation, bond lengths, charges, and angles of a periodic zincblende surface;
    \item the forces of a distorted zincblende surface;
    \item the charges, bond length and adsorption energy of \hs adsorbed on zincblende.
\end{enumerate}
Details of the training set are included in Section~F of the supplementary information.

Default ParAMS values are used for $\bm{\upsigma}$.
Individual energies, angles, distances, charges and torsions, and force groups are equally weighted in the initial training set.
The full training set, job collection and initial parameter interface are available in the supplementary information.

\section{Results and discussion}
\label{sec:results-and-discussion}

Our workflow proceeds as follows:

\begin{description}
    \item[reparameterization of the initial force field] based on the naive parameter selection (done purely for comparative purposes);
    \item[running of the sensitivity calculation] to determine a parameter ordering (entirely independent of the previous step);
    \item[rerunning the reparameterization] using 10, 20, 33 and 43 of the least and most sensitive parameters, as determined by the sensitivity analysis;
    \item[running validation tests] using the initial parameterization and some of the best parameterizations found.
\end{description}

Our aim is to show that:

\begin{enumerate}
    \item the HSIC sensitivity method correctly identifies the most sensitive parameters;
    \item reparameterizations within a lower dimensional space of very sensitive parameters can find force fields of similar quality in a shorter amount of time;
    \item reparameterizations in higher dimension run the risk of overfitting.
\end{enumerate}

\subsection{Initial optimizations}
\label{subsec:initial-optimizations}

An initial set of sixteen reparameterizations were performed using the Covariance Matrix Adaptation - Evolutionary Strategy (CMA-ES)~\cite{Hansen2001} which has been shown to work well on ReaxFF reparameterizations~\cite{Shchygol2019, Gustavo2022}.
All the optimizers were started from the initial force field values described in Section~\ref{subsec:initial-force-field} with a wide initial sampling distribution ($\sigma_0 = 0.5$).
All parameters were scaled between zero and one according to their bounds which are automatically suggested by ParAMS\@.

Optimizers were stopped if:
\begin{enumerate}
    \item within the last \num{2000} function evaluations their lowest value had not improved, and their explored values were very similar; or
    \item any of the default internal CMA-ES convergence criteria were triggered.
\end{enumerate}
The optimizers were run in parallel and the entire optimization was stopped after using a cumulative total of \num{300000} function evaluations.
These conditions are all quite strict and provided all the optimizers a very long time to find the best possible minimum.

\begin{figure}
    \centering
    \includegraphics{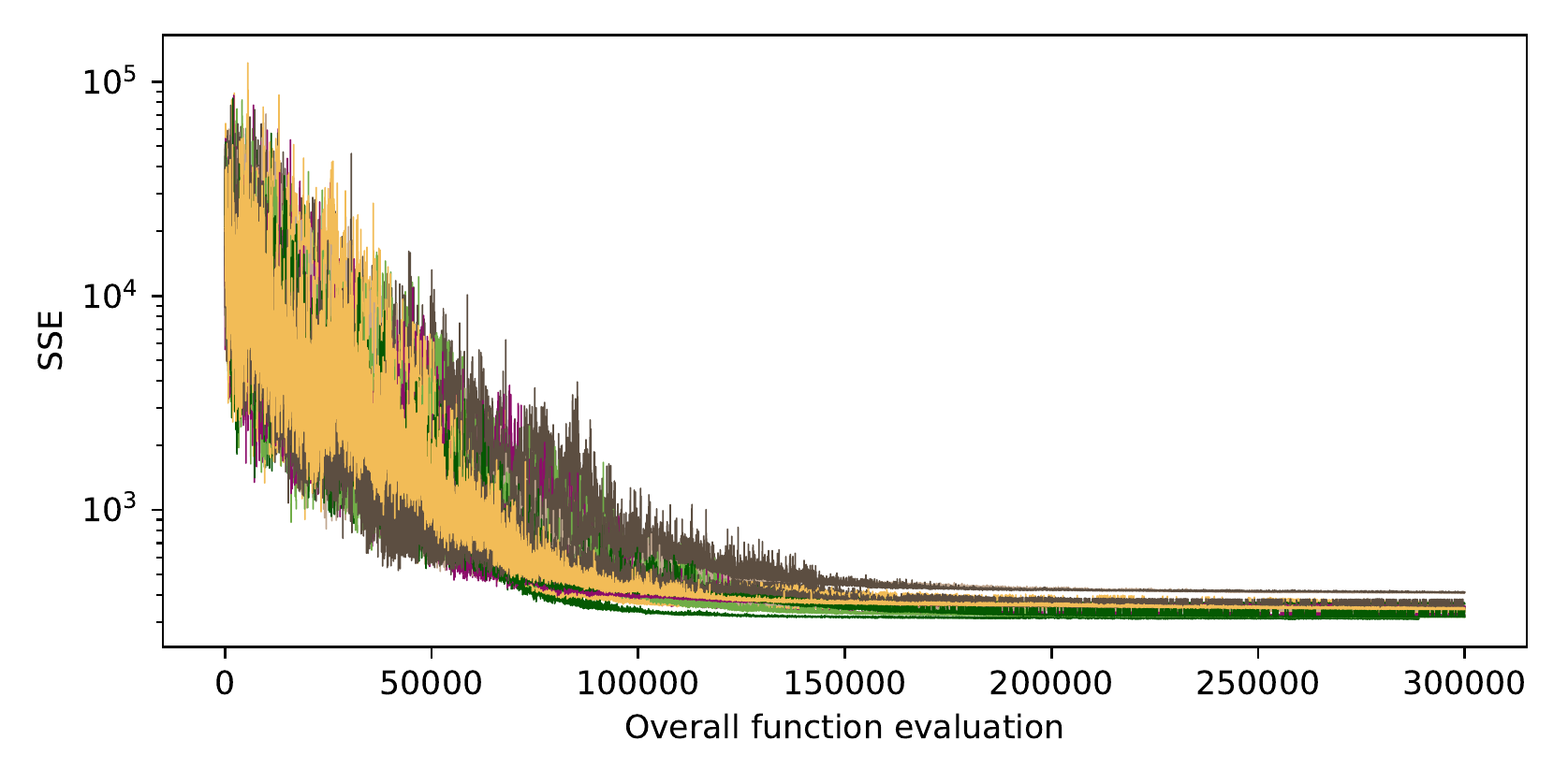}
    \caption{Optimizer trajectories for the sixteen CMA-ES reparameterizations of the initial force field using the naive parameter selection.}
    \label{fig:orig_opt}
\end{figure}

Figure~\ref{fig:orig_opt} shows the optimizer trajectories for these reparameterizations, and shows the original loss of \num{20662} being improved to between \num{410} and \num{308}.
These results will be discussed in more detail in Section~\ref{subsec:reparameterizations-with-sensitive-parameters-only} when comparing the results to other optimizations.
Detailed results for every optimizer can be found in the Section~A of the supplementary information.

\subsection{Sensitivity analysis}
\label{subsec:sensitivity-analysis}

A total of \num{10000} uniformly randomly generated samples of the original optimization space were collected for the sensitivity analysis.
The sampling procedure took a total of \SI{142}{\minute} on a \num{64} core node (AMD EPYC \num{7513} @ \SI{2.6}{\giga\hertz}).
This sampling took place once, and this same set of samples was used in all the sensitivity calculations.

\begin{figure}
    \centering
    \includegraphics{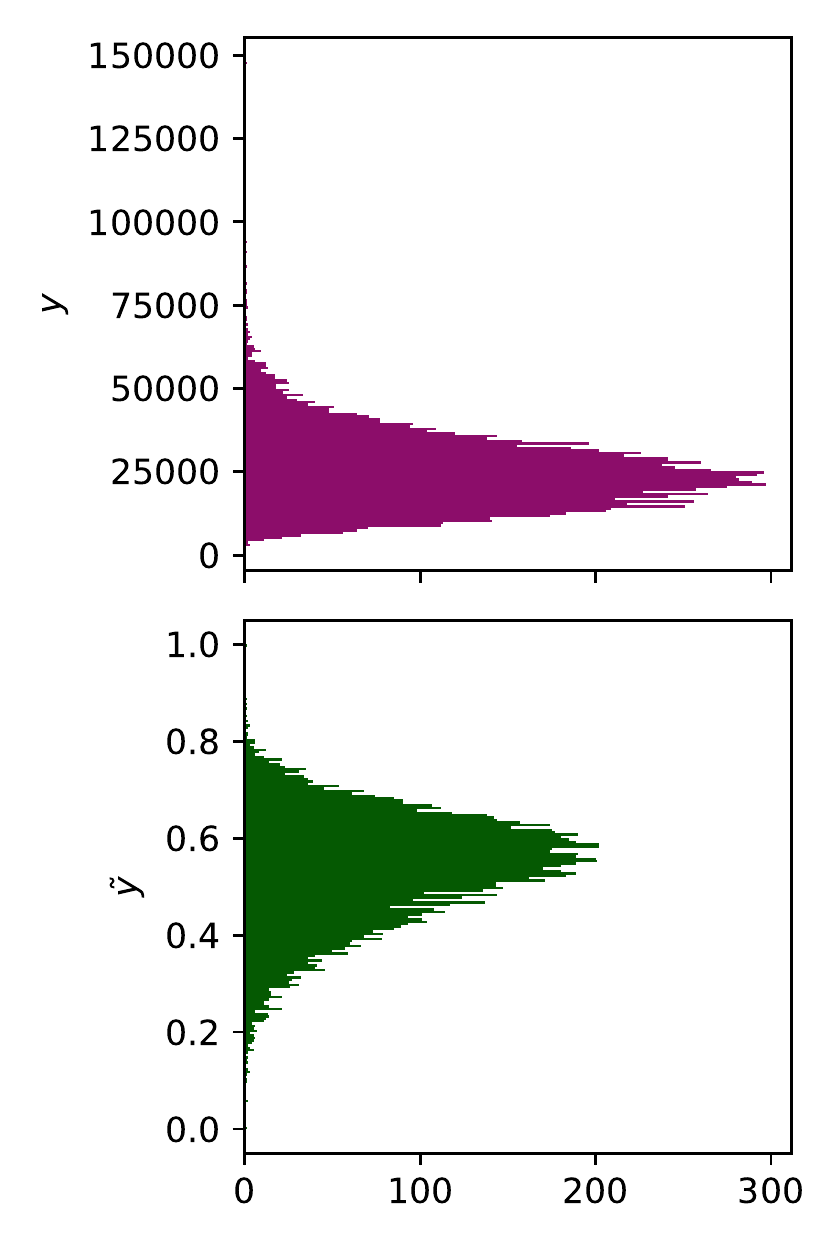}
    \caption{Distribution of loss values (top), and loss values scaled according to Equation~\ref{eq:def_ytilde} (bottom), gathered during the sampling procedure and used during the sensitivity analyses.}
    \label{fig:hist}
\end{figure}

The distribution of sampled loss values (and their transformed values, $\tilde{y}$) is shown in Figure~\ref{fig:hist};
the lowest value is \num{2651}.
Approximately half of the samples are actually worse than the initial loss value, and the set clearly does not include the good minima found through the original optimizations in Section~\ref{subsec:initial-optimizations}.
We will demonstrate that having such minima in the sample set is not a requirement for having a good estimation of parameter sensitivity.

Instead of using all of the samples at once in a single HSIC calculation, we repeated it \num{10} times using a bootstrapping method.
Each bootstrap used \num{2000} random sub-samples from the original \num{10000} sample set.
Using fewer samples per calculation significantly speeds up calculation time, and the spread of the results gives us an indication of the error.
Unless otherwise mentioned, the sensitivity values discussed below refer to the average value across the \num{10} repeats.
Algorithm~\ref{alg:sencalc} provides a pseudocode of the calculation where we have abandoned notational precision for brevity and clarity.

\begin{algorithm}
    \caption{Sensitivity calculation workflow.}
    \label{alg:sencalc}
    \begin{algorithmic}[1]
        \Require SSE, $k$, $\ell$
        \State randomly generate $\mathbf{X} \in \mathbb{R}^{\num{10000} \times 53}$
        \State $\mathbf{y} \in \mathbb{R}^{\num{10000}} \gets \mathrm{SSE}(\mathbf{X})$ \Comment{Evaluate loss}
        \State $\tilde{\mathbf{y}} \gets \tilde{y}(\mathbf{y})$ \Comment{Scale loss values}
        \For{$i=1,\dots,10$} \Comment{Repeat the calculation}
            \State $\mathbf{X}^\mathrm{sub} \gets$ \num{2000} random rows of $\mathbf{X}$
            \State $\tilde{\mathbf{y}}^\mathrm{sub} \gets$ corresponding \num{2000} loss values
            \State $\tilde{\mathbf{L}} \gets \ell(\tilde{\mathbf{y}}^\mathrm{sub})$
            \For{$\omega=1,\dots,53$} \Comment{Every parameter}
                \State $\tilde{\mathbf{K}} \gets k(\mathbf{X}^\mathrm{sub}_{\cdot \omega})$
                \State $\mathbf{H}_{i\omega} \gets \mathrm{HSIC}(\tilde{\mathbf{K}}, \tilde{\mathbf{L}})$
            \EndFor
        \EndFor
        \State $\mathbf{h} \gets$ \Call{Mean}{$\mathbf{H}_{1}, \dots, \mathbf{H}_{10}$} \Comment{Average repeats}
        \State $s_j \gets \frac{h_j}{\sum_i h_i}$ \Comment{Calculate sensitivities}
    \end{algorithmic}
\end{algorithm}

The following kernels were applied to the loss values: CG kernel ($\gamma = 0.3$), Gaussian kernel ($\sigma = 0.3$), linear kernel and threshold kernel ($\alpha = 0.4$).
In all cases a Gaussian kernel ($\sigma = 0.3$) was applied to parameter values.
Since all values are scaled between 0 and 1, a value of $\sigma = 0.3$ approximates the median distance between all samples in a uniform distribution between these values.
This has repeatedly been reported as an appropriate value~\cite{Gretton2012, Spagnol2019}.

The average time needed to perform the sensitivity calculation (regardless of loss-kernel and including the repeats) was approximately \SI{60}{\second}.
The optimizations require several days to complete, thus the inclusion of these sampling and sensitivity steps in a reparameterization workflow would not be the limiting step and, as will be shown below, come with several advantages.

We note that we also repeated the calculations using a sample pool of \num{30000} samples, and \num{10000} sub-samples per calculation.
These results produce tighter distributions between the repeats at the cost of more time, but the average sensitivity values are essentially the same.
In other words, they were not worth the effort and are not presented below.

The scaling of the sensitivity calculation is $\mathcal{O}(r d n^2)$ where $r$ are the number of repeats, $d$ is the number of parameters, and $n$ is the number of sub-samples used per calculation.
Increasing the size of the matrices has the largest impact on the calculation time, so being able to get robust parameter orderings from repeated calculations using small sub-samples is very advantageous.

\begin{figure}
    \centering
    \includegraphics{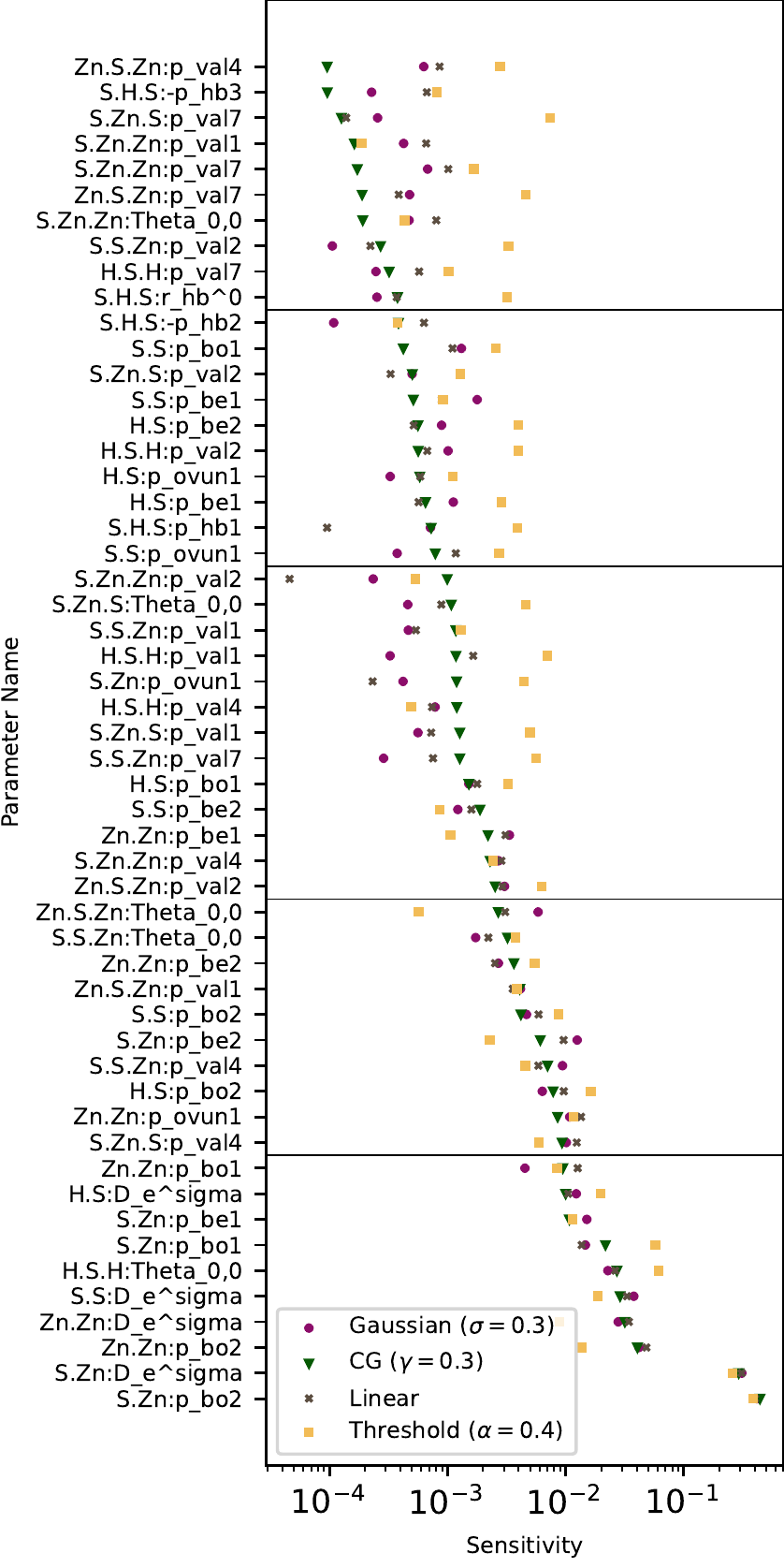}
    \caption{Average sensitivity values (over ten runs each) determined for each parameter using different kernels applied to the loss values. Parameters are ordered by the results of the CG kernel which is used in the later reparameterizations. Black lines group the \numlist{10;20;33;43} most sensitive parameters.}
    \label{fig:compare-sensitivities}
\end{figure}

Figure~\ref{fig:compare-sensitivities} compares the sensitivity values determined by each calculation for each parameter.
Despite the large differences between loss-kernels, the ordering of parameters is very robust for the most sensitive parameters.
Significant deviations only appear for the least sensitive parameters which are all close to one another numerically (note the logarithmic axis).
This demonstrates that the HSIC approach we have taken is robust, and not overly dependent on the choice of loss-kernel or hyperparameters.

The loss-kernel which struggles the most with differentiating sensitivities is the threshold kernel.
We believe that this is because the kernel is discontinuous and provides very little information to the calculation because values are toggled to zero or one.
Our investigation of this kernel was motivated by the work of \citet{Spagnol2019}.
They demonstrated the necessity of this kernel to focus the calculation on small loss values (near the minima of interest), and negate the effects of large order-of-magnitude differences.
As discussed in Section~\ref{subsec:kernels}, our implementation always takes the logarithm of the loss values and then scales the result;
it appears to address the order-of-magnitude problem.
Random sampling will also generally not contain very good minima (true in this case, as discussed above), making the second advantage of the threshold kernel less important.
The continuity provided by the other kernels seems to make converging the sensitivity values for the parameters easier.

We have also explored the effect of using different kernel parameters in Section~B of the supplementary information.
This has a minor effect on the sensitivity results, provided a reasonable kernel parameter value is selected so that the kernel values are not all very close to zero or one.

Given the similarities in orderings obtained by the various loss-kernels we will continue discussing only the results obtained when using the CG kernel ($\gamma=0.3$).
This is a somewhat arbitrary choice, but it can be argued that it has the best theoretical foundation by emphasizing the sensitivities for low minima.

\begin{figure}
    \centering
    \includegraphics{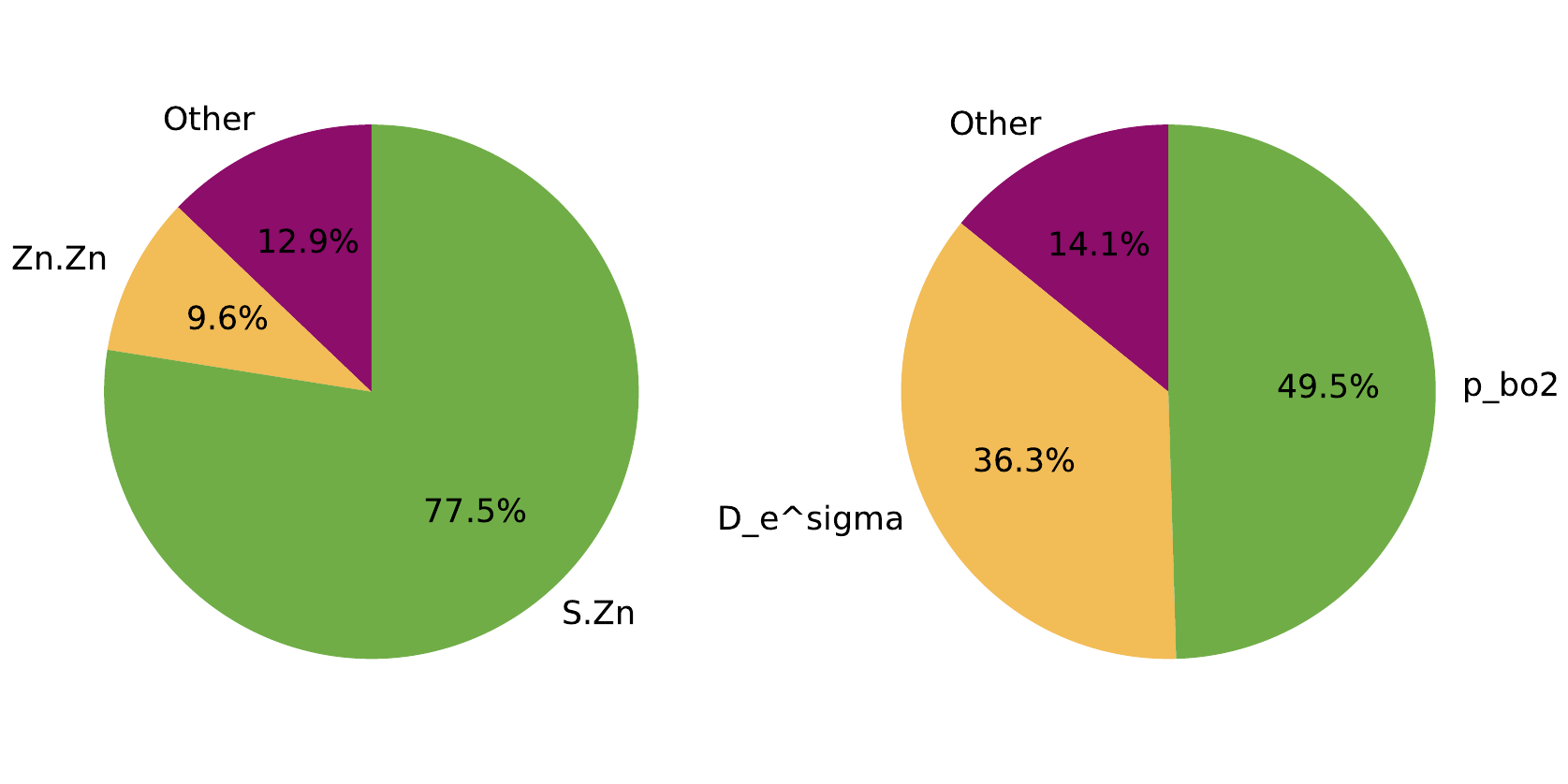}
    \caption{Grouped sensitivities as determined by the HSIC calculation using the CG kernel. Grouped by parameter group (left) and parameter name (right).}
    \label{fig:origpies}
\end{figure}

Figure~\ref{fig:origpies} shows the sensitivities obtained grouped by a) parameter group, and b) parameter name.
For example, `S.Zn' refers to all the bond parameters associated with sulphur-zinc bonds, and \texttt{p{\_}bo2} refers to all the \texttt{p{\_}bo2} bond parameters regardless of atoms involved.
\texttt{p{\_}bo2} appears in the calculation of $\sigma$-bond orders, and is an exponential term.
\texttt{D{\_}e\textasciicircum sigma} is a linear parameter which is multiplied by the $\sigma$-bond orders to determine the bond energy contribution to the overall potential (see Equation~\ref{eq:bond_energy}).

The identification of these $\sigma$-bond parameters, particularly for sulphur-zinc and zinc-zinc bonds are quite reasonable given the composition of the training set.
The degree to which the sensitivity is dominated by only these parameters, however, may be surprising.
Nevertheless, the results seem intuitive.

\subsection{Reparameterizations with sensitive parameters only}
\label{subsec:reparameterizations-with-sensitive-parameters-only}

Based on the sensitivities determined above, we ran reconfigured reparameterizations using: the most sensitive \numlist{10;20;33;43} parameters, and the least sensitive \numlist{10;20;33;43} parameters.
The most-sensitive groupings are shown in Figure~\ref{fig:compare-sensitivities}.
We made these selections so that the set of \num{10} most sensitive parameters, and the set of \num{43} least sensitive parameters are complementary;
together they account for the original \num{53} parameters.
Similarly, for other combinations.
Each of the reparameterizations was conducted in the same way as the original described in Section~\ref{subsec:initial-optimizations}.

In a more practical setting we do not advocate blindly activating only the most sensitive $n$ parameters;
we do so here only to avoid introducing human decision-making into the results.
In practice, one should use the sensitivities as a guide to better understand how the training data effects the loss function, and use some human intuition when selecting parameters.

For the sake of clarity we introduce the following nomenclature to refer to the different optimizations:
\begin{description}
    \item[\orig] reparameterizations using the original 53 active parameters;
    \item[\most{\#}] reparameterizations using the most sensitive \# parameters;
    \item[\least{\#}] reparameterizations using the least sensitive \# parameters.
\end{description}

\begin{figure}
    \centering
    \includegraphics{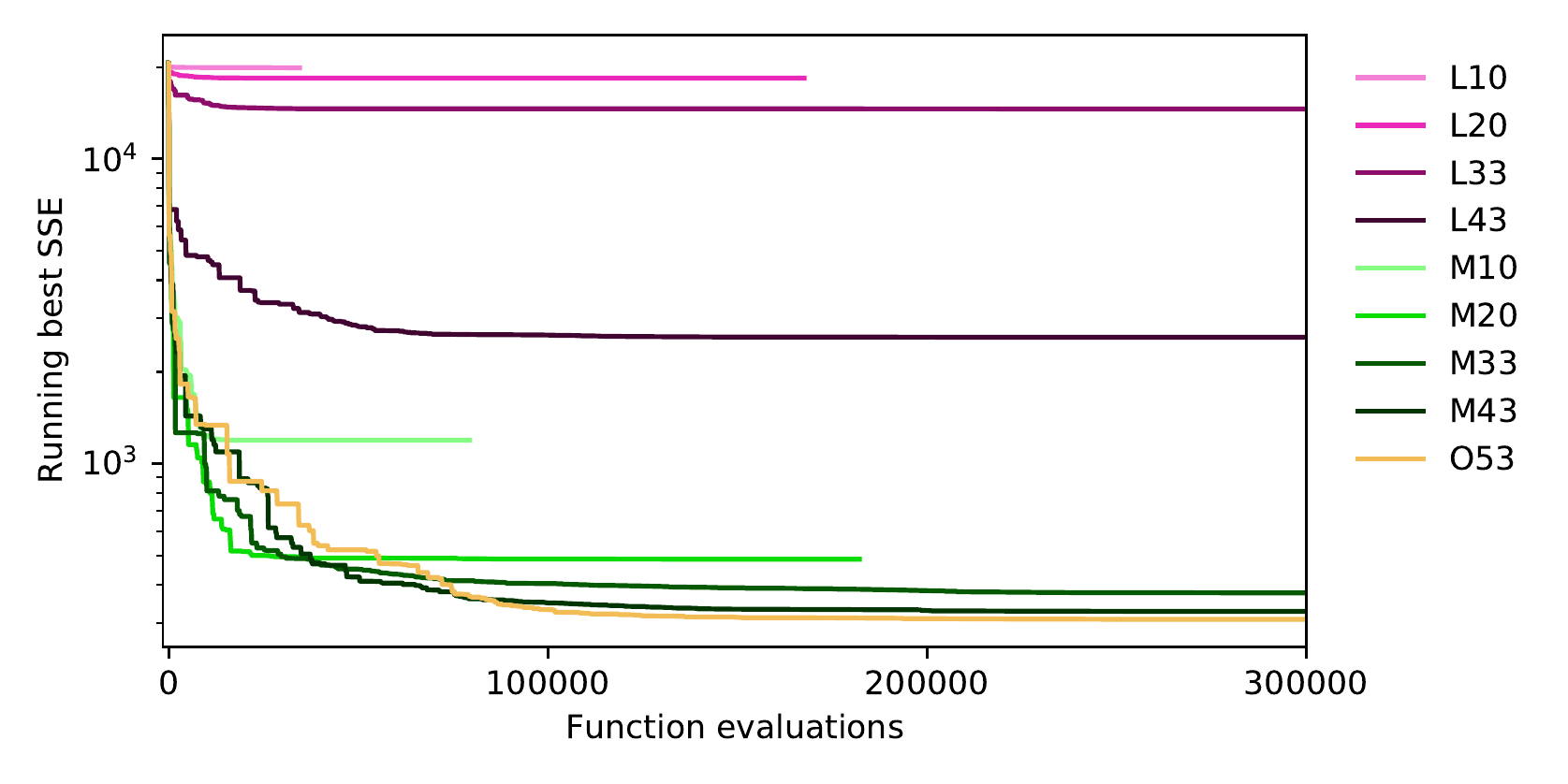}
    \caption{Running best value seen by any of the sixteen parallel optimizers for each of the optimization configurations.}
    \label{fig:optimizations}
\end{figure}

Figure~\ref{fig:optimizations} shows the running best loss value seen by any of the sixteen parallel optimizers for each of the optimization configurations (Section~A of the supplementary information contains the results for individual optimizers).
Table~\ref{tab:err_removed} shows the difference between the lowest loss value and initial loss value as a fraction of the initial loss value.
In other words, the percentage of the initial loss which was `removed' during the optimization.
Unsurprisingly, \orig produces the lowest loss because it had the most degrees-of-freedom.
Any reduction in the number of active parameters can be expected to worsen achievable loss values.
Indeed, all the other optimizations find worse minima than the original.
However, the difference between using the most sensitive and least sensitive parameters is marked.

\begin{table}
    \centering
    \caption{Fraction of the initial loss value which removed during optimization.}
    \label{tab:err_removed}
    \begin{tabular}{lSS}
        \toprule
           & {Most Sensitive (\%)} & {Least Sensitive (\%)} \\
        \midrule
        10 & 94.2 & 3.61 \\
        20 & 97.7 & 10.9 \\
        33 & 98.2 & 29.4 \\
        43 & 98.4 & 87.4 \\
        53 & 98.5 &  \\
        \bottomrule
    \end{tabular}
\end{table}

In all cases using the most sensitive parameters allows us to remove \SI{93}[>]{\percent} of the original error.
In fact, some \most{33} and \most{43} optimizers find better minima than some \orig optimizers.
In contrast, using the same number of least-sensitive parameters produces very poor optimizations which are unable to meaningfully reduce the loss value.
Especially noteworthy is that \most{10} is able to locate better minima than its complement \least{43}.
Similarly, \least{10} is only able to reduce the loss by \SI{4}{\percent} compared to \SI{94}{\percent} achieved by \most{10}.

It is also important to note that \most{10} and \most{20} are able to converge long before the \num{300000} function evaluation limit, and use approximately one-third and two-thirds of the time used by \orig.
This represents significant time savings as \orig took approximately two days to complete, more than the time required to run the sensitivity analysis and \most{10} or \most{20}.

These results show that our proposed HSIC sensitivity method is able to quickly identify the most sensitive parameters for reparameterization, and that a parameter selection guided by it can produce good minima in a shorter time.

\subsection{Force field comparison and validation}
\label{subsec:force-field-comparison-and-validation}

In order to verify the quality of the new force fields, and demonstrate the risks of overfitting, we perform several validation tests using the initial force field, and the force fields with lowest error produced by \orig, \most{10}, and \most{20} across the sixteen optimizations each one performed.

We consider:
\begin{enumerate}
    \item the error on the training set items;
    \item an MD simulation of \hs adsorption on zincblende and wurtzite slabs;
    \item an MD simulation of zincblende bulk crystal;
    \item the adsorption energy of an \hs molecule on a \zplane zincblende surface; and
    \item the surface energies of \zplane zincblende and \wplane wurtzite.
\end{enumerate}

We compare the results to reference DFT calculations, as well as literature values.
For calculations involving crystal surfaces, we use the \zplane face of zincblende and \wplane face of wurtzite ZnS as they have been found to be the most energetically favorable~\cite{Zhang2003, Hamad2002}.
Slabs were always constructed from pre-optimized lattice parameters calculated with BAND using the PBEsol/DZP level of theory.

\subsubsection{Training set errors}

\begin{figure}
    \centering
    \includegraphics{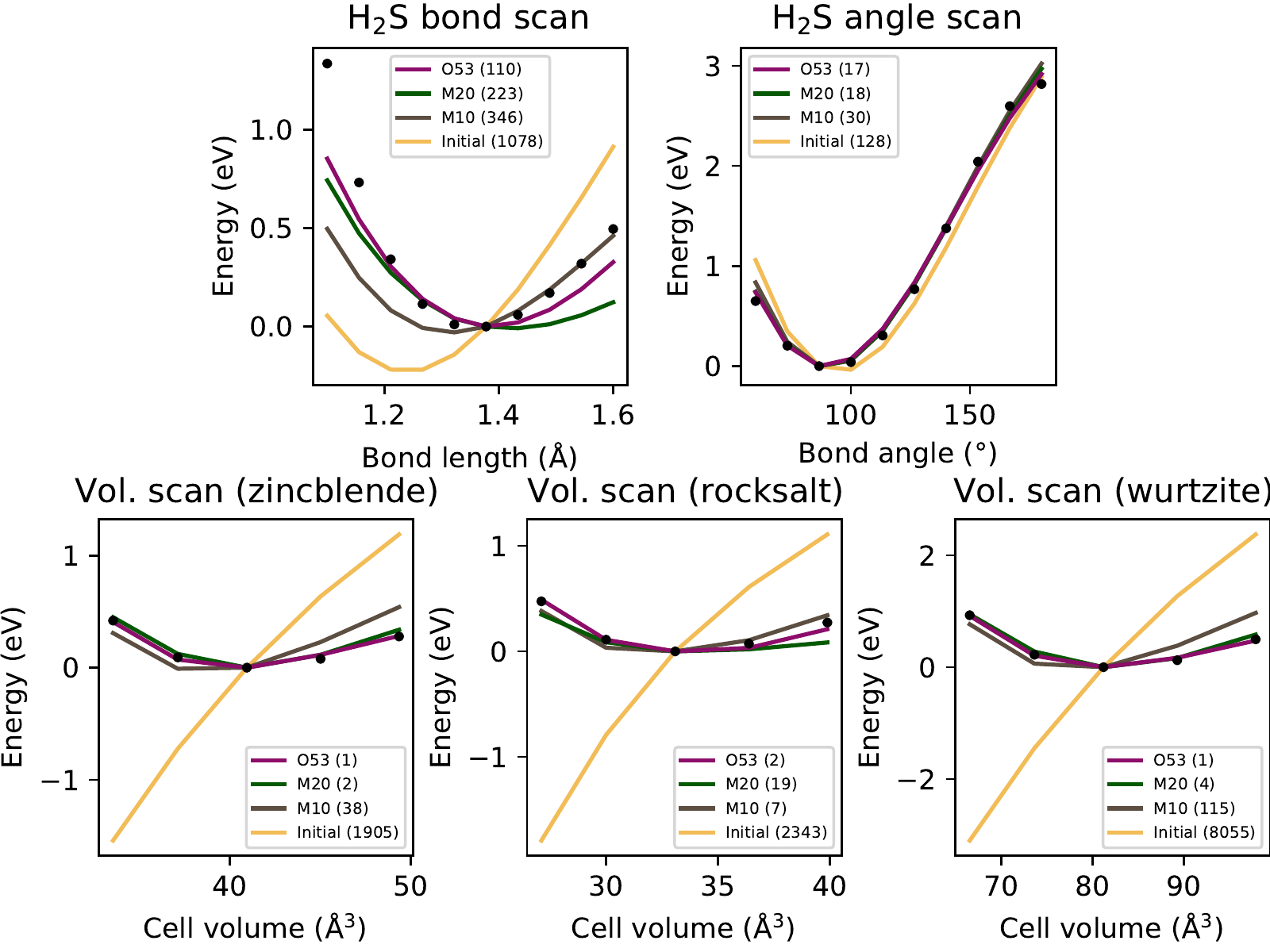}
    \caption{Training set items which form part of various PES scans. Reference values shown as black dots and loss values for each force field are shown in parentheses in the legends.}
    \label{fig:pes_scans}
\end{figure}

\begin{figure}
    \centering
    \includegraphics{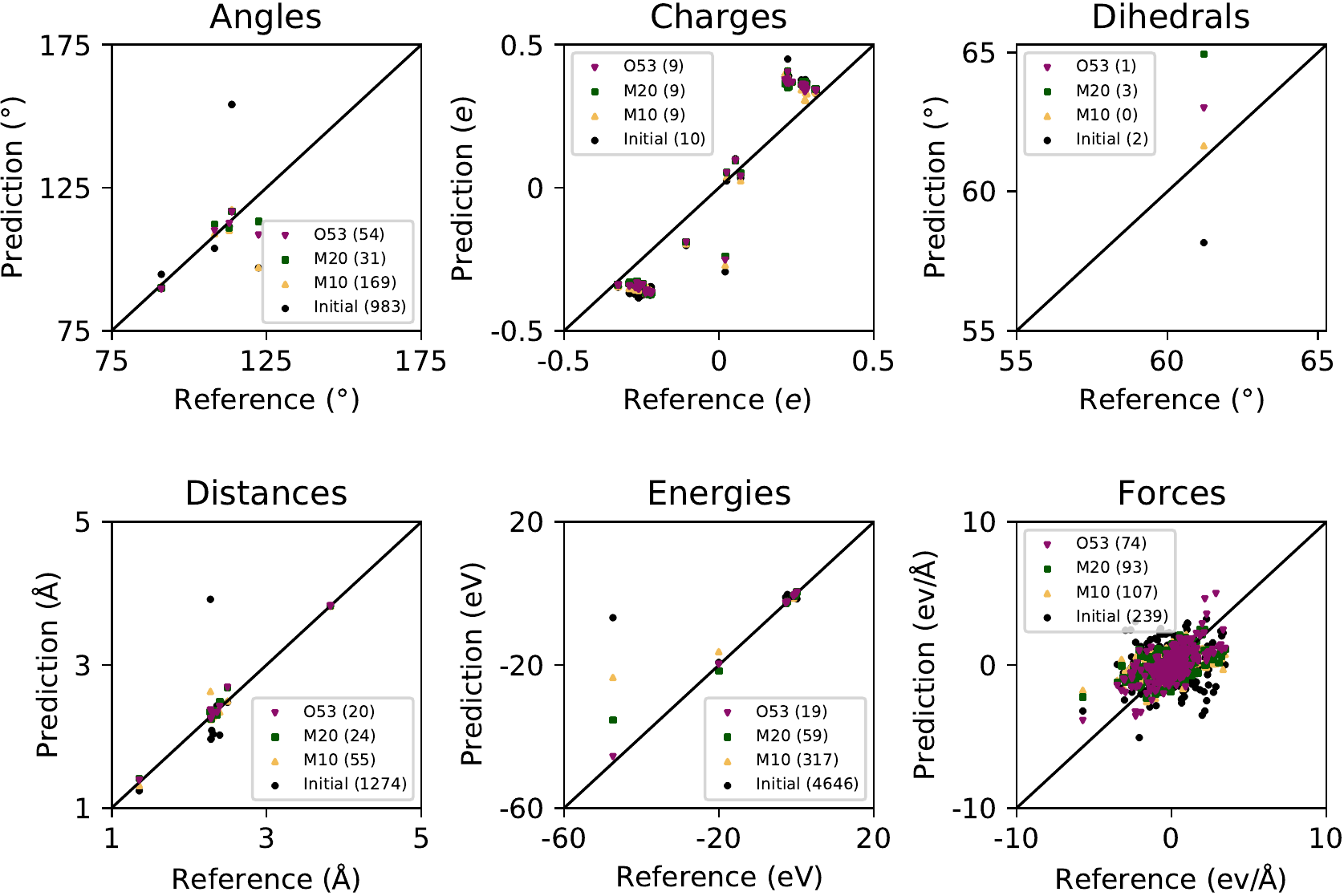}
    \caption{Comparison of reference to predicted values for each of the training set items and force fields, grouped by item type. Loss values for a group and force field are included in parentheses in the figure legends.}
    \label{fig:scatter_plots}
\end{figure}

We start our analysis by decomposing the overall loss values for each of the force fields into their individual contributions.
Figures~\ref{fig:pes_scans} and~\ref{fig:scatter_plots} compare force field predictions to reference values for each of the training set items.

Figure~\ref{fig:pes_scans} shows training set items which can be grouped together into PES scans along some coordinate.
With the exception of the \hs angle scan, the original force field performs extremely poorly for these items, and in all cases the optimized force fields produce significantly improved results.
The \orig force field is the most accurate for all the scans, but \most{20} is generally quite competitive and has similar errors.
The one exception to this is the H-S bond scan in the \hs molecule which \most{20} did not replicate well.

Figure~\ref{fig:scatter_plots} shows the remaining training set items where we compare reference values to predictions made by the selected force fields.
Although angles and forces were improved, significant errors remained in most cases and a large fraction of the original forces error remains in the final fields.
Energies, which form a large part of the overall loss value, actually appear generally well-fitted and most of the error comes from a single item;
this item is the energy difference between a distorted and regular \zplane zincblende surface.
We can also see that charges were effectively unimproved during the reparameterization.
This is because charge related parameters were not activated in our parameter selection.
Interatomic distances were the most improved class of training set items.
This can be explained by the fact that the original sulphur related parameters were actually trained for oxygen, thus the initial force field bond lengths were too short to accommodate the larger sulfur atoms and had to be lengthened.

In most cases the \most{10} errors are substantially worse then the \most{20} ones.
On the other hand, \most{20} errors are often very similar to (or in the case of angles) better than the \orig errors.
In comparison to \orig, \most{20} performs worst for forces and energies.
However, as mentioned above, the bulk of the energy error is concentrated in one item, and no force field appears to predict forces very well.
From these observations, there does not appear to be a strong signal that \orig is a significantly better force field than \most{20}.

\subsubsection{MD simulations}
\label{subsubsec:md-simulations}

\paragraph{Adsorption to crystal surfaces}

Our first validation test is an MD simulation of the adsorption of \hs molecules on ZnS slabs.
We are guided by literature in constructing a realistic simulation scenario.
\citet{Zhang2003} showed computationally and experimentally that cubic ZnS with a particle size smaller than $\sim$\SI{7}{\nano\meter} is not stable at room temperature and can easily convert to the hexagonal polymorph.
Experimentally, \citet{Dloczik2001} were able to produce stable ZnS columns with wall thicknesses between \SIrange{10}{30}{\nano\meter} where both cubic and wurtzite phases were detected.
\citet{Qi2014} later studied the adsorption of \hs on ZnS surfaces at room temperature.

Using these sources, we create MD simulations for both morphologies using slabs $\sim$\SI{10}{\nano\meter} thick.
The temperature is maintained at \SI{298}{\kelvin}, \num{100} \hs molecules are randomly positioned above and below the slabs, and the simulations are conducted for \num{300000} time steps of \SI{0.25}{\femto\second} each.

\begin{figure}
    \centering
    \subcaptionbox{Initial force field}{\includegraphics[width=0.49\textwidth]{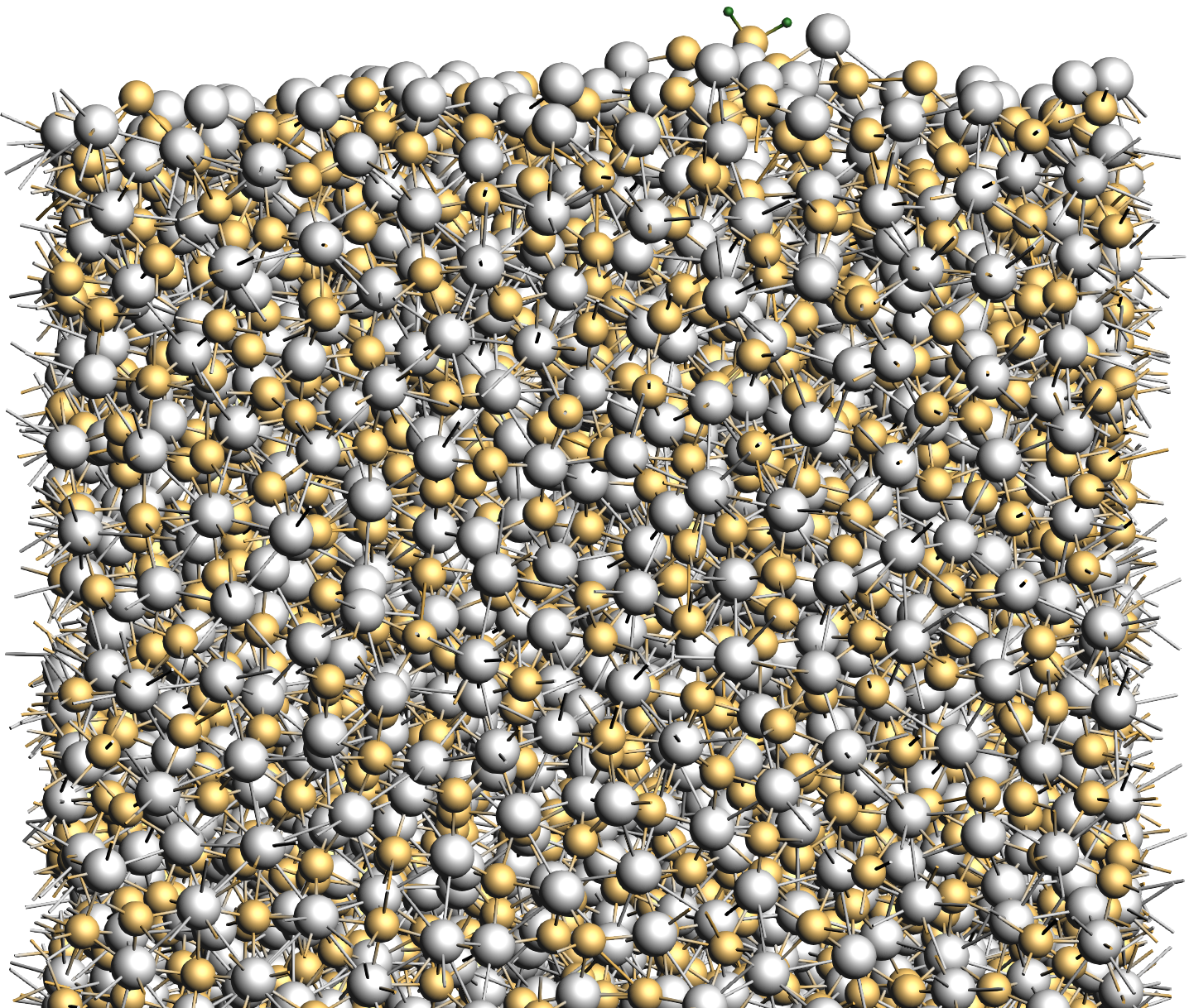}}
    \subcaptionbox{\orig}{\includegraphics[width=0.49\textwidth]{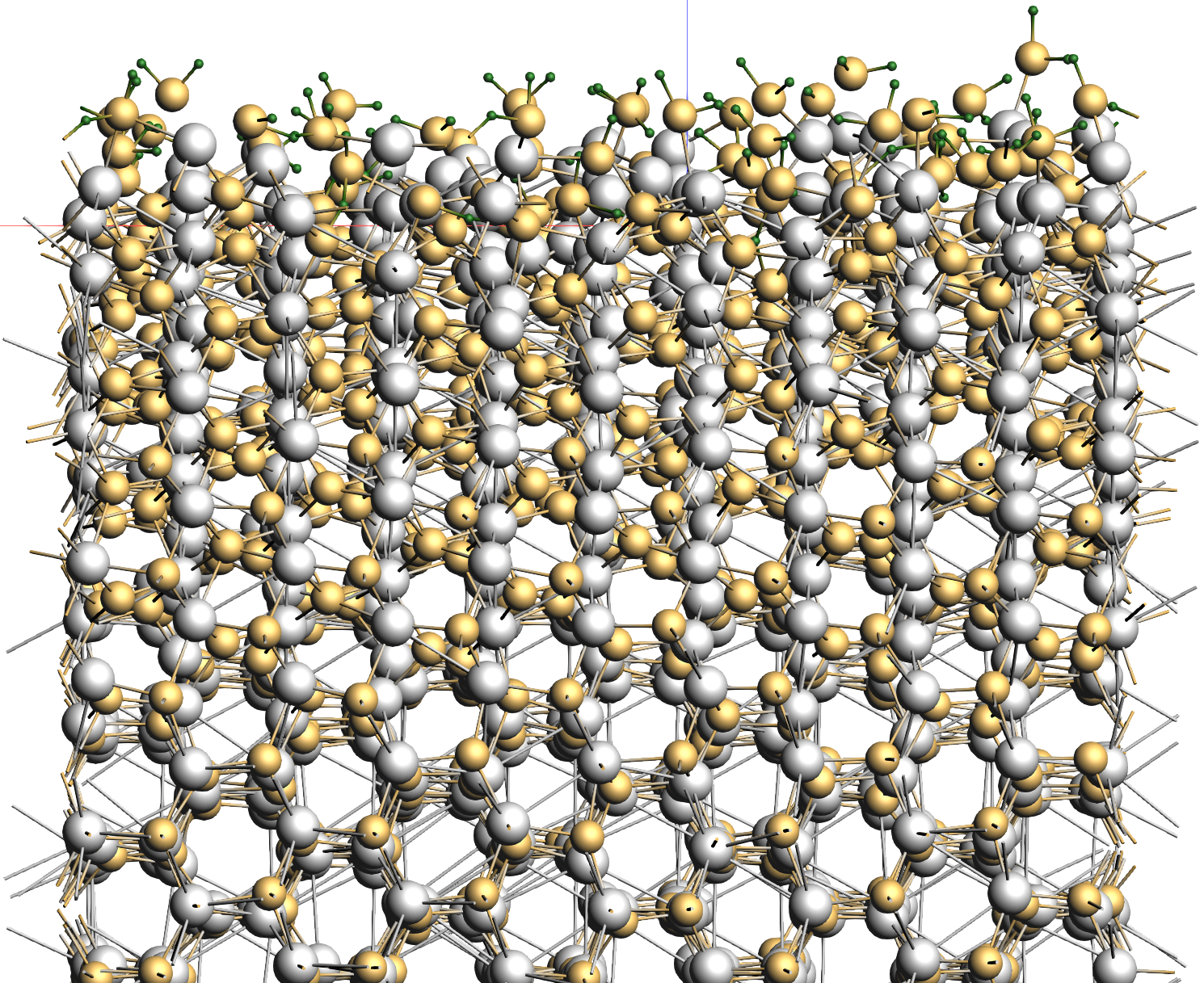}}
    \subcaptionbox{\most{10}}{\includegraphics[width=0.49\textwidth]{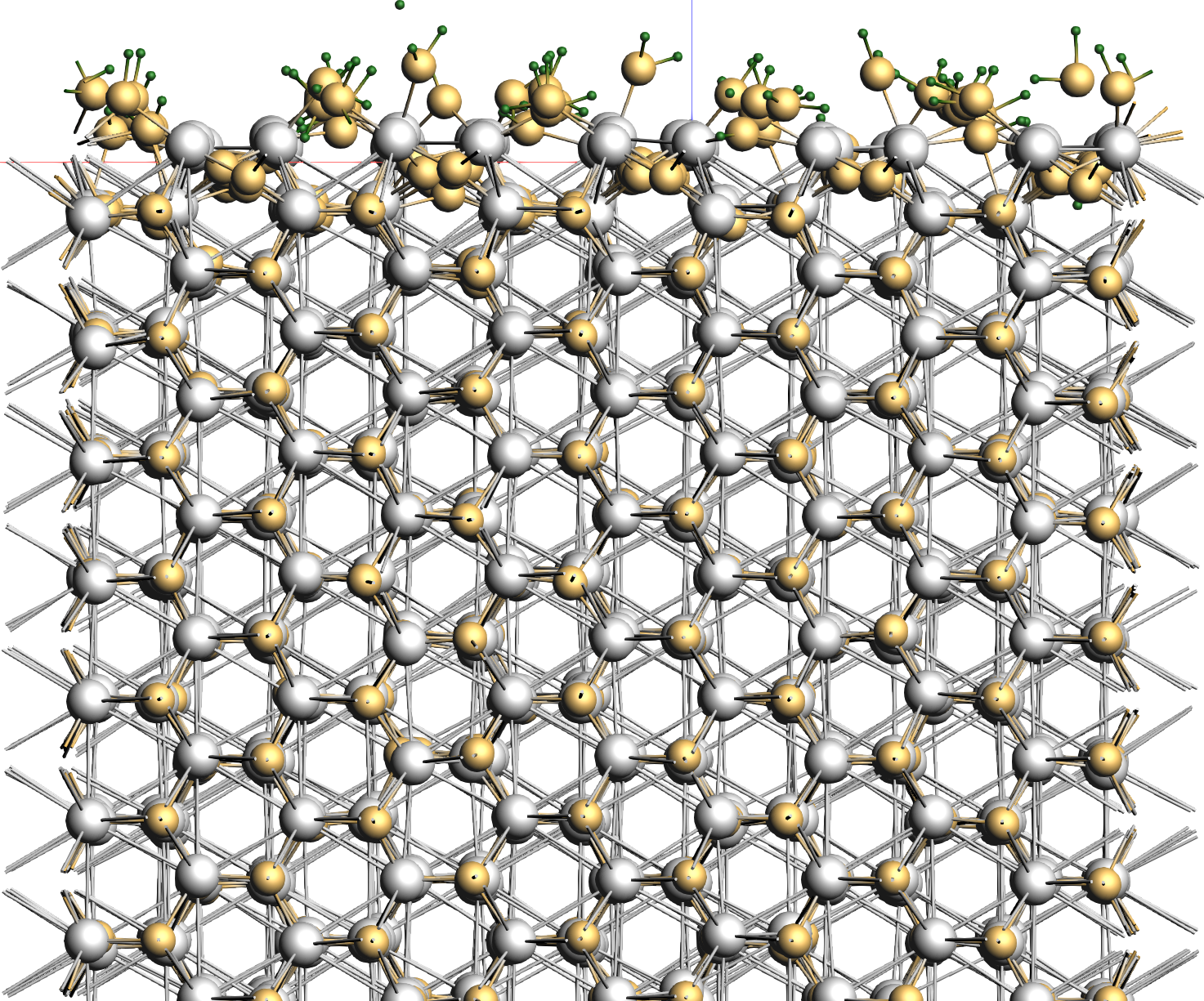}}
    \subcaptionbox{\most{20}}{\includegraphics[width=0.49\textwidth]{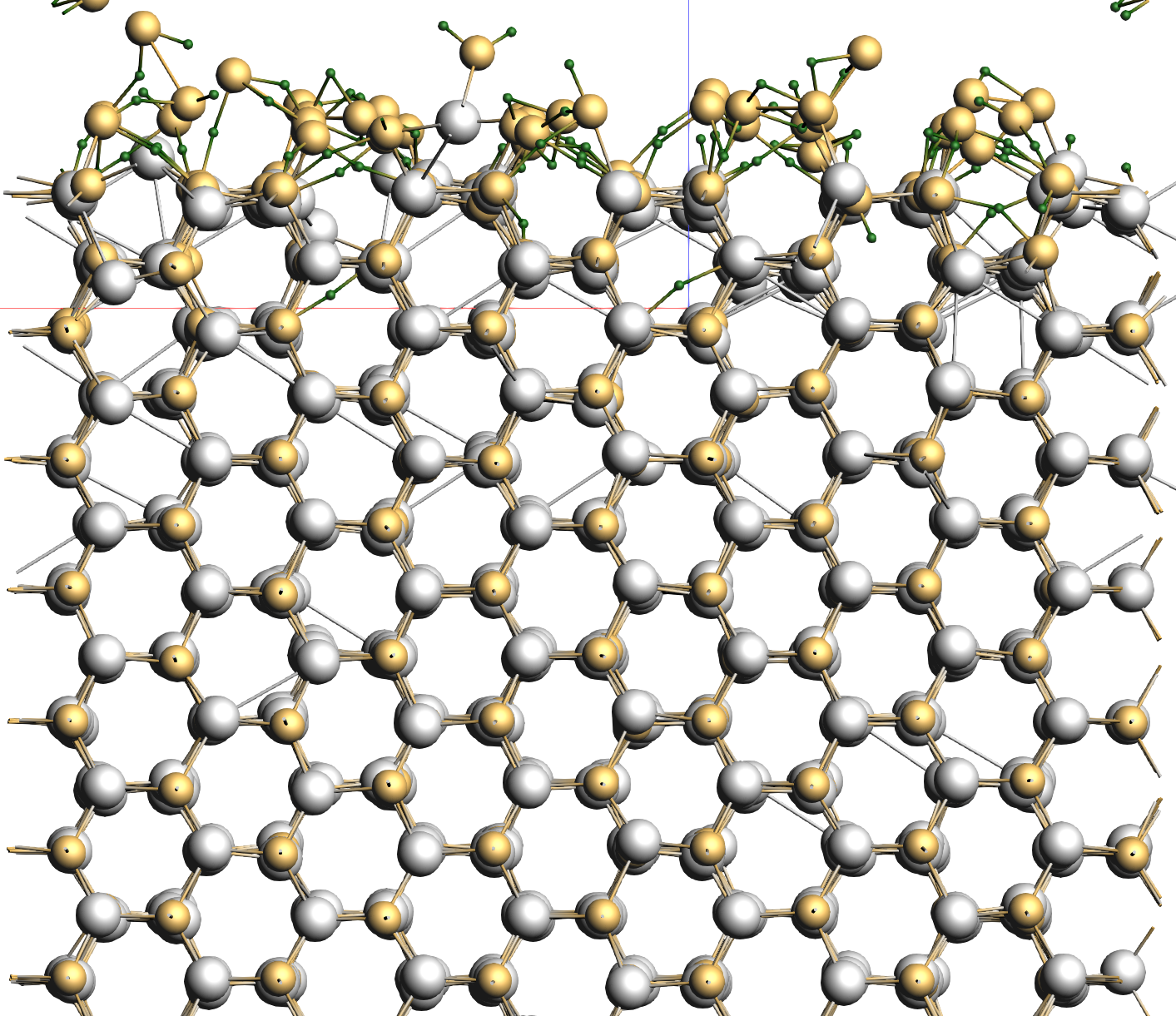}}
    \caption{One surface in the final frames of \SI{75}{\pico\second} MD simulations of \num{100} \hs gas molecules above and below a \SI{10}{\nano\meter} thick wurtzite slab with \wplane surfaces using a) the initial, b) \orig, c) \most{10}, and d) \most{20} force fields. Sulfur atoms in yellow, zinc in silver, hydrogen in green.}
    \label{fig:md_wurtzite}
\end{figure}

\begin{figure}
    \centering
    \subcaptionbox{Initial force field}{\includegraphics[width=0.49\textwidth]{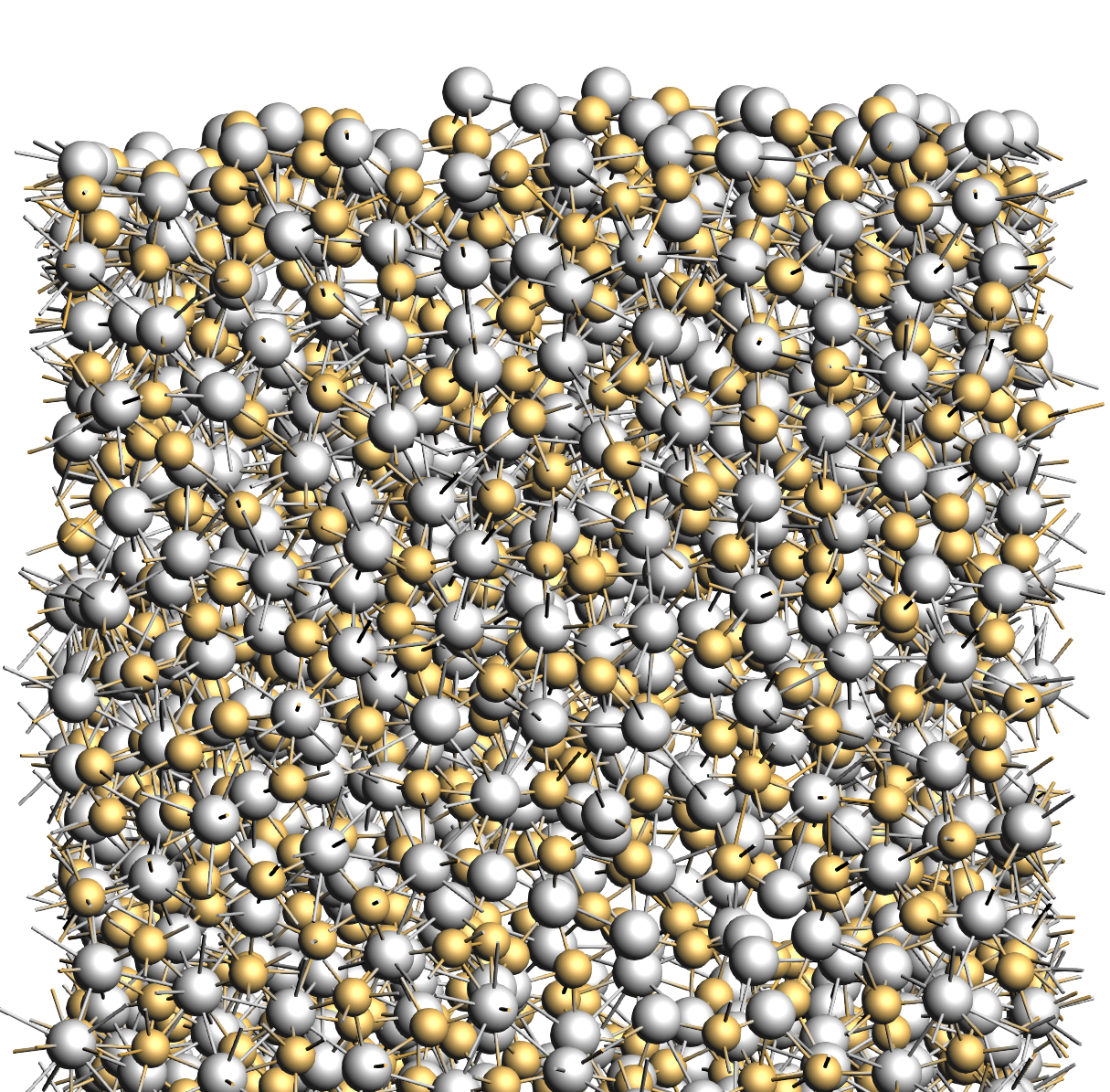}}
    \subcaptionbox{\orig}{\includegraphics[width=0.49\textwidth]{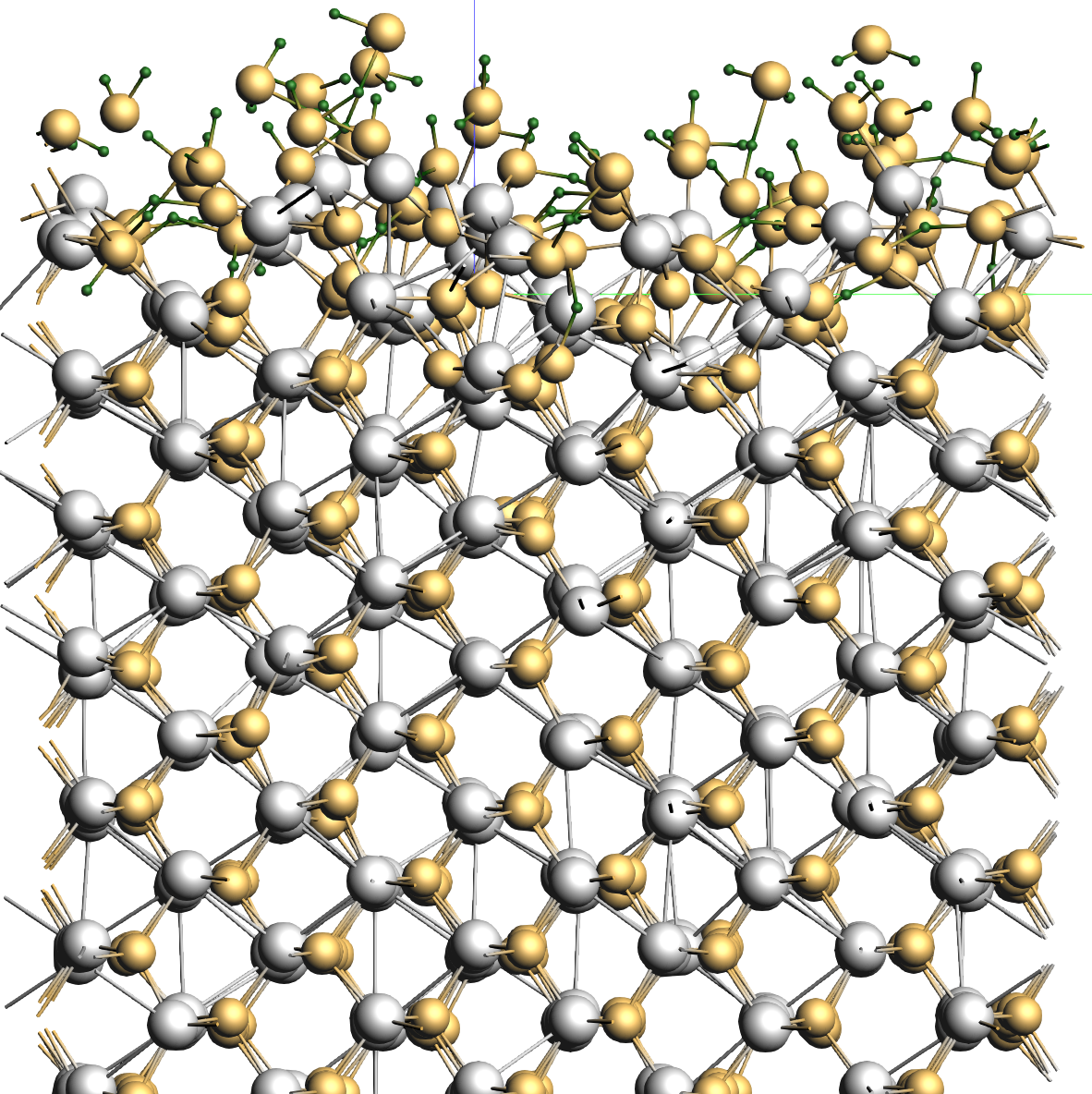}}
    \subcaptionbox{\most{10}}{\includegraphics[width=0.49\textwidth]{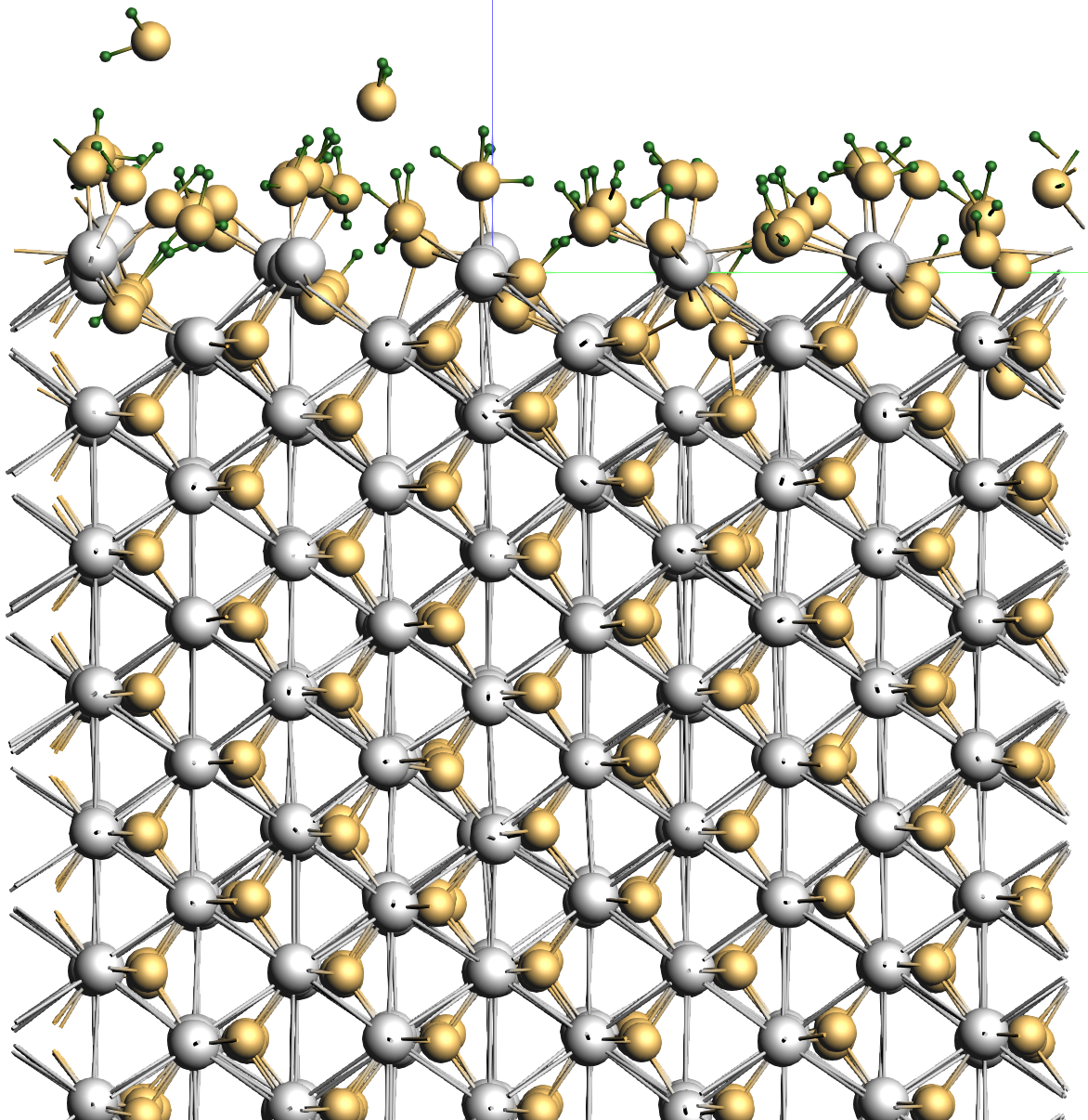}}
    \subcaptionbox{\most{20}}{\includegraphics[width=0.49\textwidth]{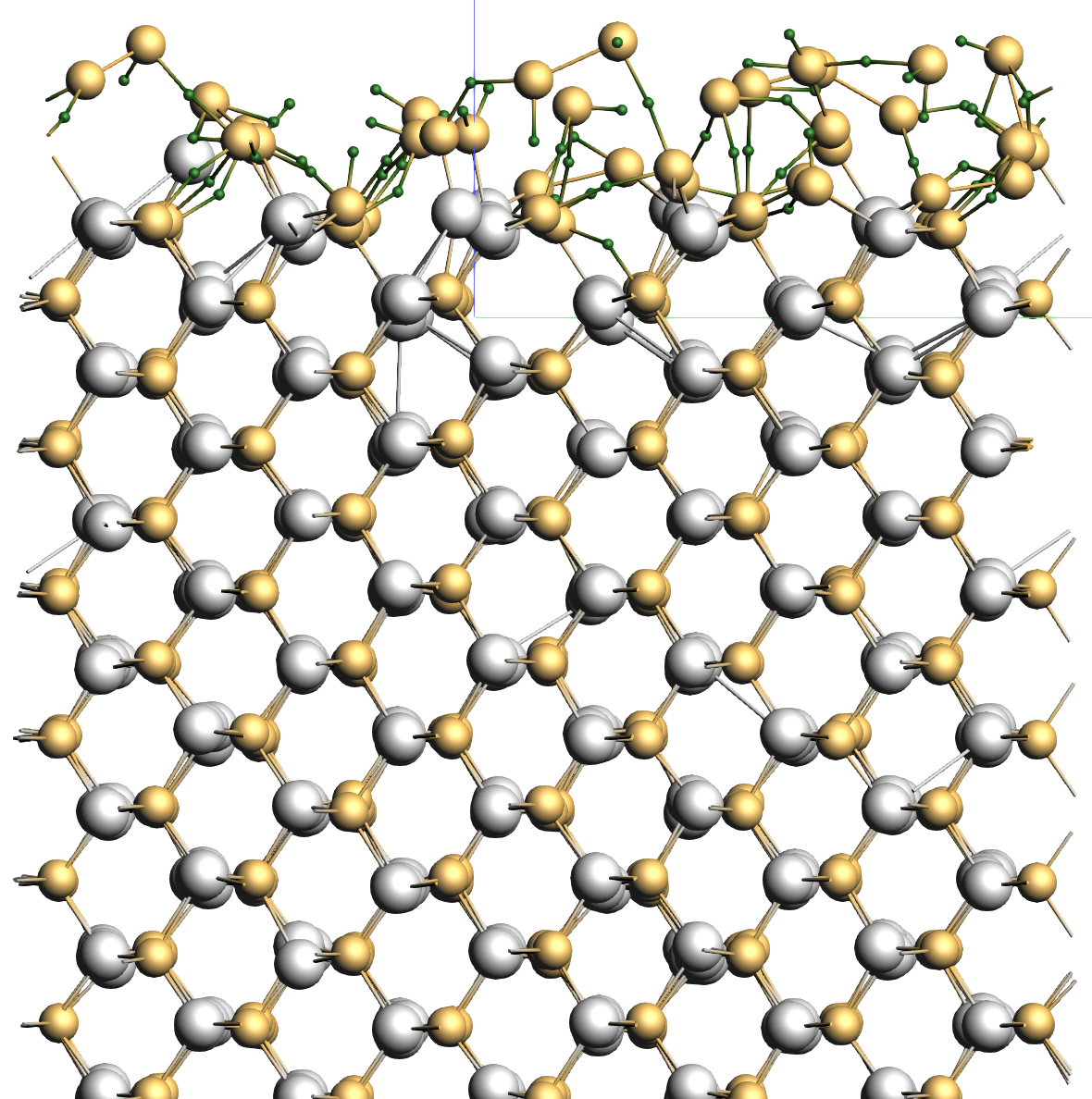}}
    \caption{One surface in the final frames of \SI{75}{\pico\second} MD simulations of \num{100} \hs gas molecules above and below a \SI{10}{\nano\meter} thick zincblende slab with \zplane surfaces using a) the initial, b) \orig, c) \most{10}, and d) \most{20} force fields. Sulfur atoms in yellow, zinc in silver, hydrogen in green.}
    \label{fig:md_zincblende}
\end{figure}

Figures~\ref{fig:md_wurtzite} and~\ref{fig:md_zincblende} show the final surface geometries for the wurtzite and zincblende simulations respectively.
Unsurprisingly, the initial force field performs poorly as the slabs immediately collapse.
The \most{10} and \most{20} simulations, however, perform better.
The slabs maintain their crystalline shapes, and adsorb the \hs molecules to the slab surfaces.

The \orig simulations do not behave in the same way.
These slabs struggle to maintain their surfaces.
In the wurtzite case the surface layers have become almost amorphous, although the bulk crystal is relatively well-maintained, the extent of the deformation extends several layers deep.
The zincblende surface is more ordered, but it is not as rigidly maintained as \most{10} or \most{20}, and it appears that some \hs molecules are being chemically absorbed into the structure.

As discussed above, literature suggests that these slabs should be stable rigid solids under these conditions, making the behavior of the \orig force field surprising.
We will discuss the reasons for this behavior in Section~\ref{subsubsec:adsorption-and-surface-energies}.

\paragraph{Bulk crystal}

The above simulation provides a qualitative indication that the lower loss value of the \orig force field does not guarantee better performance.
However, we would like to demonstrate quantitatively that \most{20} parameterizations produce more robust results.

To do this we ran a second MD simulation of a bulk zincblende crystal (\num{64} atoms total) at \SI{500}{\kelvin} for \SI{10}{\pico\second} using a time step of \SI{0.5}{\femto\second}.
Two simulations were run: one used the best force fields produced by each of the sixteen \most{20} optimizers, and the other used the best force fields produced by each of the sixteen \orig optimizers.
In each simulation the force fields were used as a committee.
In other words, each force field was applied to the same crystal geometries at each time step.
The forces of these results were then averaged to update the atomic positions for the next time step.

\begin{figure}[htpb]
    \centering
    \includegraphics{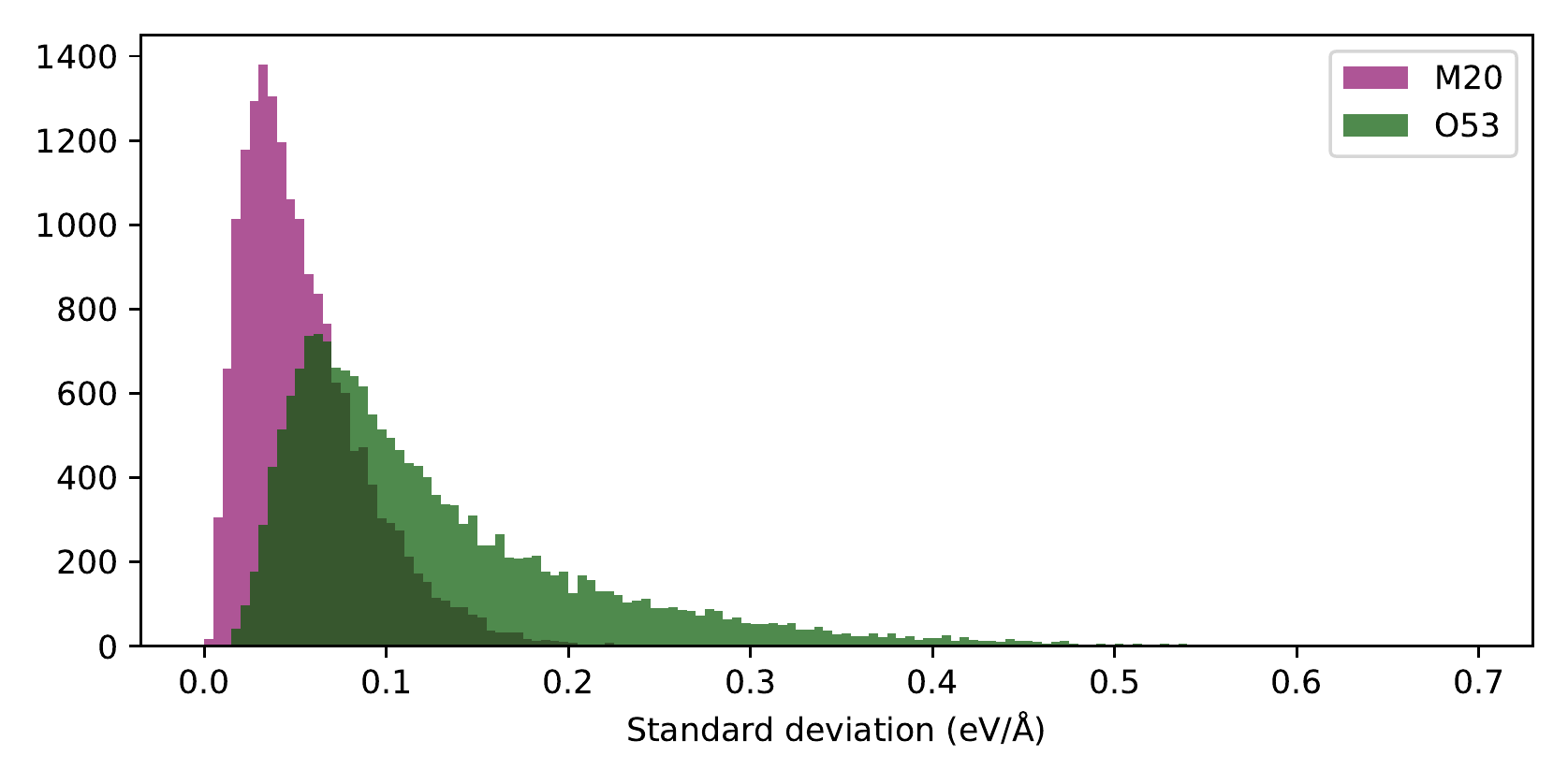}
    \caption{Standard deviation on the forces felt by all atoms, in all directions, at all time steps in bulk simulations of zincblende crystal using committee simulations of the sixteen \most{20} and \orig force fields.}
    \label{fig:hybrid}
\end{figure}

Figure~\ref{fig:hybrid} shows the standard deviation of the forces across the sixteen force fields.
We are not interested in the results at particular times, for particular atoms, or in particular directions, because the forces always average out to zero.
Thus, the histograms simply show all the standard deviations regardless of the above distinctions.
The sixteen \most{20} force fields show significantly more agreement than the \orig ones.
This suggests that the lower dimensional optimization is much more robust, and produces force fields with a tighter distribution of results.
The higher dimensional optimization has many more minima in which optimizers might be trapped.
Since these extra minima are created by the presence of insensitive parameters, they are more likely to represent overfittings than genuine attractors driven by important parameters.
Interestingly, the favorable spread of results seen here for \most{20}, could not be assumed from the spread in loss values amongst its sixteen force fields which was \num{98} compared to \num{33} for \orig.

\subsubsection{Adsorption and surface energies}
\label{subsubsec:adsorption-and-surface-energies}

Our more quantitative validation tests are:
\begin{enumerate}
    \item the calculation of the adsorption energy ($E_\mathrm{ads}$) of an \hs molecule on the \zplane surface of zincblende; and
    \item surface energy ($E_\mathrm{surf}$) calculations for the \zplane surface of zincblende and \wplane surface of wurtzite.
\end{enumerate}
Details of these calculations are included in Section~C and Section~D of the supplementary information, respectively.

\begin{figure}
    \centering
    \includegraphics{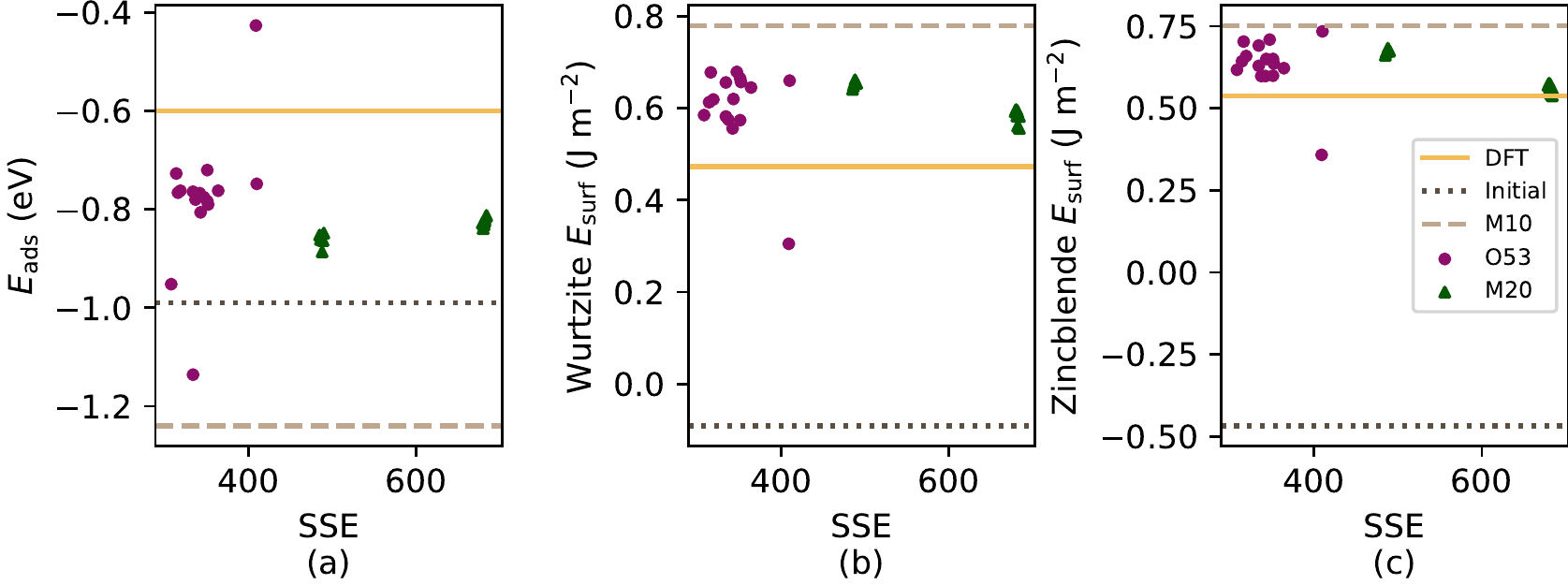}
    \caption{a) \hs adsorption energy, b) wurtzite \wplane surface energy, and c) zincblende \zplane surface energy versus SSE for the best force fields found by each of the sixteen optimizers run in the \orig and \most{20} reparameterizations. Reference DFT results shown with horizontal yellow line.}
    \label{fig:sse-validation}
\end{figure}

In this case we use the best force fields from every optimizer in the \orig and \most{20} sets to analyse the distribution of results.
We do not consider all sixteen \most{10} results as they perform poorly.
The results are shown in Figure~\ref{fig:sse-validation} as a function of the force fields' associated SSE\@.
Detailed results for individual optimizers can be found in Section~E of the supplementary information.
In these plots, we have also included DFT, initial force field, and best \most{10} force field results for comparison.
The initial and \most{10} results are not shown with their corresponding loss values as they are too large and would make the plots illegible.
In the discussions that follow, we will continue to use `best' to refer to the force field with the lowest loss value (SSE), it does not mean the force field with the most accurate validation result.

\paragraph{Adsorption energy}

We begin our comments by considering the adsorption energy results.
None of the force fields are very accurate in this regard in comparison to our DFT reference, however the best \most{20} force field is more accurate than that from \orig.
Interestingly, the initial force field produces a better prediction than \most{10} and is only slightly worse than the best \orig force field.
More broadly, however, all the \most{20} force fields predict adsorption energies in a narrow range from \SIrange{-0.89}{-0.81}{eV}, which is much more reproducible than the \orig results which range from \SIrange{-1.14}{-0.43}{eV}.

\begin{figure}
    \centering
    \subcaptionbox{}[\linewidth]{\includegraphics[width=\linewidth]{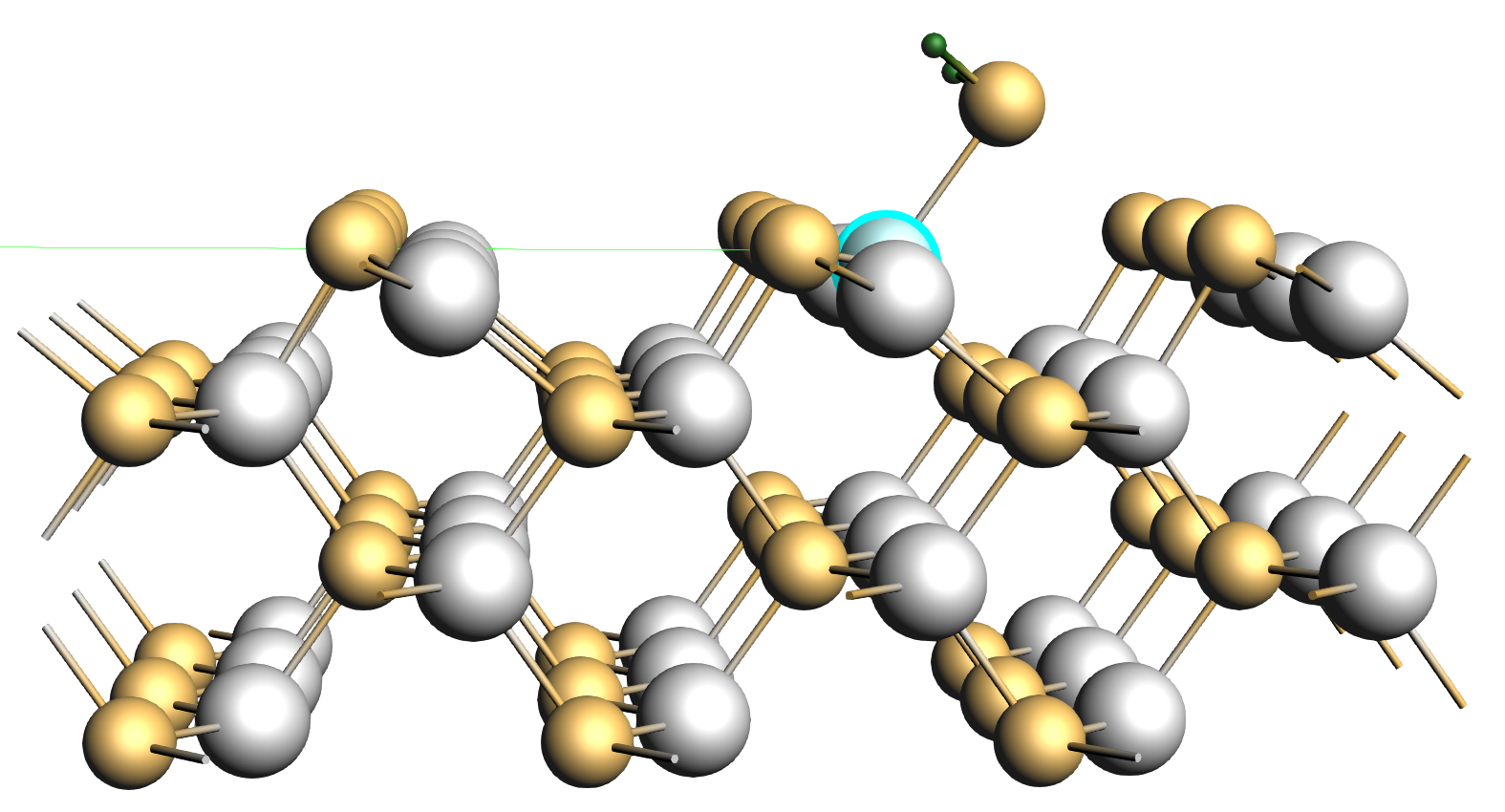}}
    \subcaptionbox{}[\linewidth]{\includegraphics[width=\linewidth]{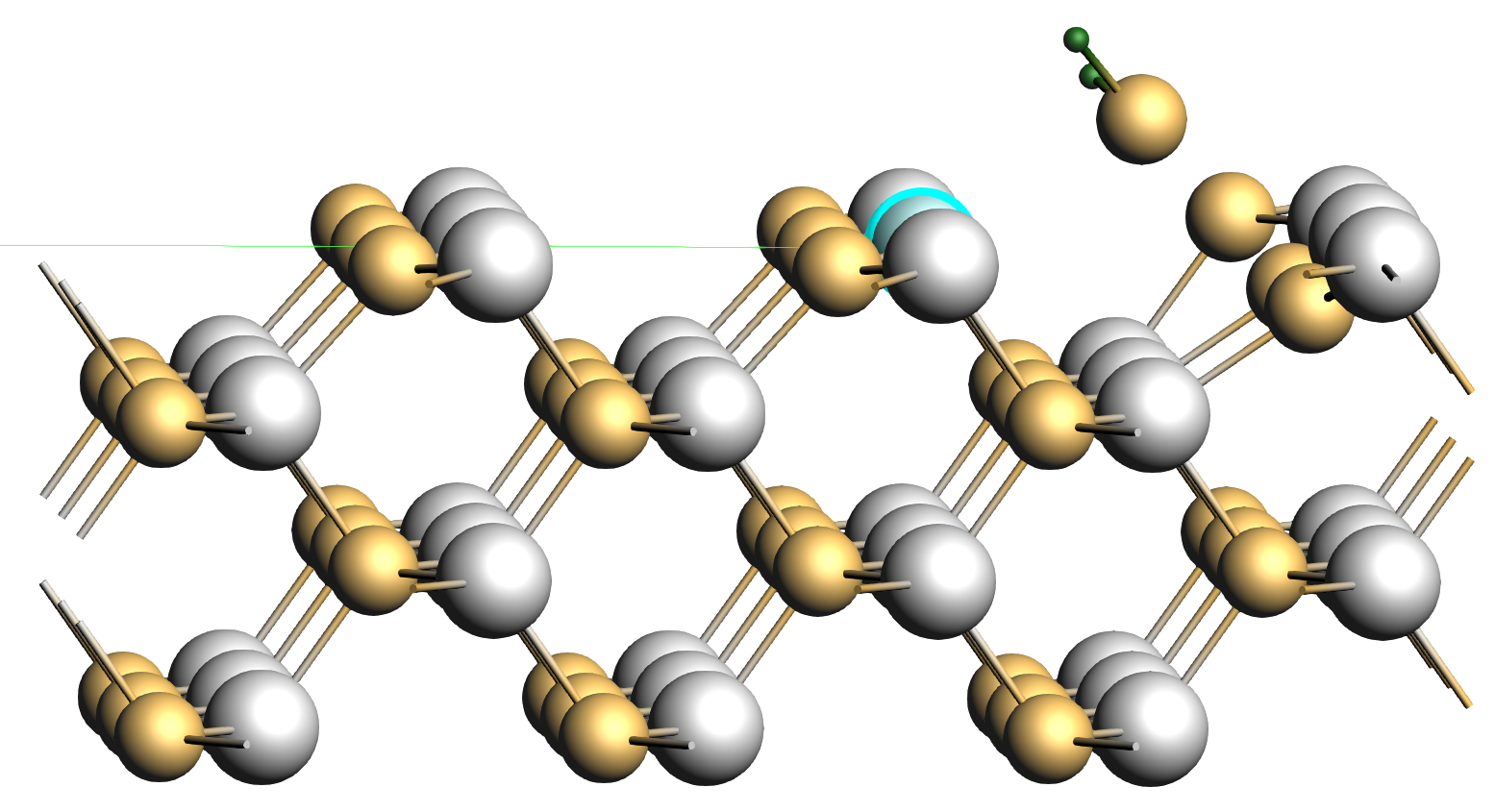}}
    \caption{Final frame of a geometry optimization of an \hs molecule adsorbed on a \zplane zincblende surface using a) DFT and b) the best \orig force field. The molecule is originally positioned near the zinc atom in blue.}
    \label{fig:adsorption}
\end{figure}

This variability might be explained by the odd behavior seen in several \orig geometry optimizations.
Figure~\ref{fig:adsorption} shows the final frames the adsorbed molecule geometry optimizations using DFT and the best \orig force field.
The \hs molecule is initially placed close to the zinc atom highlighted in blue, and settles into this position using DFT\@.
However, using the \orig force field, the \hs molecule moves away from this atom towards a row of sulfur atoms which are, in turn, repelled by it.
The geometry optimization is converged, which means that this desorbed position with deformed surface is more energetically favorable than an adsorbed position on zinc---which is incorrect.
It is possible that the strange surface effects seen in Section~\ref{subsubsec:md-simulations} may be associated with this.

\paragraph{Surface energy}

If we consider the surface energy results, the distributions of the \orig force fields are consistently larger than those of the \most{20} results.
On average, the \orig results are similarly accurate to \most{20} for the wurtzite surface.
For the zincblende surface, one of the \most{20} clusters is actually quite accurate.

Overall, none of the force fields performed generally well across all the validation tests, but they are improvements to the initial force field.
If more accuracy is desired, then more attention needs to be given to the composition of the training set, however, this is not the focus of this article.

\paragraph{Correlations with loss values}

The most interesting conclusions from these results, come from comparing them to the force fields' associated SSE\@.
All sixteen \orig optimizers found SSE values tightly clustered between \num{308} and \num{410}, much lower and less variable than the \most{20} optimizers that are clustered around two values between \num{485} and \num{684}.
If one only considers the loss values, then one might expect that predictions made by the \orig force fields will be similarly more precise than those of \most{20}, however, we see that this is not the case.
\orig validation results are significantly more variable than those of \most{20} and, on average, seem to be only slightly more accurate.

Most crucially, the high variability of the \orig validations are not correlated with the SSE\@.
In other words, a better prediction of the training set items is not correlated with a more accurate prediction of the validation tests.
We believe that these facts, and the unusual behavior of the \orig MD simulations in Section~\ref{subsubsec:md-simulations}, can be attributed to overfitting.
Interestingly, the \most{20} results seem to be \textit{inversely} correlated with the loss values, i.e., a higher loss leads to a more accurate validation result.
This might suggest that \most{20} is also seeing some overfitting effects, however, there is not enough data to verify this because all optimizers fall near one of only two SSE values.

Overfitting is a common occurrence during ReaxFF reparameterization~\cite{Muller2016, Shchygol2019, Hubin2016} and becomes more likely as more parameters are allowed to change.
These validation tests highlight the dangers of activating too many dimensions during a reparameterization as more degrees of freedom create opportunities for overfitting.
We present two dangers which occur when too many parameters are used.

First, it is important to emphasize the difference between a parameter's \textit{sensitivity} to some loss function (which is a function of a user-generated training set), and its true \textit{importance} in the potential function.
If a training set does not include enough items to capture the importance of a parameter, then it will register as insensitive.
If an important, but insensitive, parameter is activated during reparameterization then the optimizer will change it quasi-randomly.
This can result in force fields with very low loss values, that perform very poorly in production runs;
sensitive parameters have been set correctly, but other important ones have been changed incorrectly.

A second danger is the fact that parameters can have a compensatory effect.
If many parameters are active, then it is possible that multiple parameter sets can satisfy the training data, but the potential will perform poorly during simulation.
By using too many parameters the optimizers have an opportunity to move insensitive ones to reduce the loss rather than identifying the `correct' values for a small number of appropriate parameters.

In our example, the \num{53} dimension reparameterizations seem to have activated too many dimensions.
Although they produced the lowest overall loss values, they did not achieve the lowest loss for every class of items, nor did they always make the most accurate or precise predictions in our validation tests.
In the MD simulations we see odd behavior which, along with the precision problems discussed previously, we believe to be evidence of overfitting.

Conversely, the reparameterizations using only the \num{10} most sensitive parameters seem to not have had enough degrees of freedom.
These force fields produce significant training set errors compared to \orig and \most{20}, and perform poorly in the validation tests.

Using \num{20} of the most sensitive parameters produces good quality force fields which perform better than the \orig force fields in some tests, despite having marginally worse training set losses.
The results of the validation tests were also consistently more precise than those of \orig, and of comparable accuracy.
The MD simulation aligned with expectations, and the force fields were produced in a shorter time than that required for the original optimization.

\section{Outlook}
\label{sec:outlook}

We believe that there are several areas available which warrant further study.
First, the training set used here could be used as a starting point to produce a better quality force field for Zn/S/H\@.
This would require a significant expansion of the training set, and involve all of the well-known and thorny issues associated with training set design~\cite{Muller2016}.

Second, a parameter's sensitivity is a strong function of the range in which it is allowed to vary.
In this work, we have not needed to give this much consideration since the ParAMS package provides recommended ranges.
However, it is not certain that these are always appropriate.
A formalisation of appropriate ranges, or a robust mechanism to determine them would go a long way to ensuring that sensitivities are being appropriately determined.

Third, we have identified sampling as the limiting step of the sensitivity procedure.
It may be interesting to investigate the extraction of samples from optimizer trajectories.
However, since the HSIC requires uniformly distributed random samples, one would need to extract samples carefully from the trajectories.
One possibility is the Kennard-Stone algorithm~\cite{Kennard1969}, but this is slow and sequential.
Nevertheless, `closing the loop' and allowing users to extract sensitivities from optimization results seems like an appealing prospect.

\section{Conclusions}\label{sec:conclusions}

As an introductory work, we have demonstrated that an HSIC sensitivity analysis applied to a ReaxFF reparameterization can successfully identify the most sensitive parameters.
We have further shown that using only the most sensitive parameters during optimization leads to faster convergence and a reduced chance of overfitting.
Even qualitatively, the use of such a sensitivity analysis can provide valuable insights for the user into the composition of the training set.
Overall, we believe that the HSIC sensitivity analysis is a useful, robust, and easy to use tool which has the potential to greatly aid in the reparameterization of ReaxFF force fields.

\section*{Author Contributions}
\label{sec:author-contributions}

MG designed and implemented the sensitivity test, ran the test work, and wrote this article.
MH designed the training set and guided some of the validation tests.
TV provided supervision, revisions, guidance, and advice.

\begin{acknowledgement}
The authors thank the Flemish Supercomputing Centre (VSC), funded by Ghent University, FWO and the Flemish Government for use of their computational resources.

Funding for the project was provided by the European Union's Horizon 2020 research and innovation program under grant agreement No.\ 814143. TV and MG are also supported by the Research Board of Ghent University (BOF) under award No.\ 01IT2322.
\end{acknowledgement}

\begin{suppinfo}
The following items are included in the Supplementary Information:
\begin{description}
    \item[\texttt{si.pdf}] full optimization results, calculation methodologies for adsorption and surface energies, full adsorption and surface energy results, training set composition;
    \item[\texttt{job\_collection.yaml}] training set geometries and job calculation settings;
    \item[\texttt{engine\_collection.yaml}] task settings for the input geometries;
    \item[\texttt{training\_set.yaml}] reference values, weights, and sigma values for training data;
    \item[\texttt{initial\_parameters.yaml}] original ReaxFF parameter values and ranges (\orig parameters set as active).
\end{description}

The YAML files are human-readable and used by ParAMS~\cite{params}.

This information is available free of charge via the Internet at \url{http://pubs.acs.org}.
\end{suppinfo}

\bibliography{bibliography}

\end{document}


\maketitle
\thispagestyle{fancy}
\pagenumbering{gobble}
\tableofcontents
\appendix

\newpage
\pagenumbering{arabic}

\section{Full optimization results}
\label{sec:full-optimization-results}

\begin{table}[htpb]
    \centering
    \caption{Lowest SSE loss value found by every parallel optimizer for every set of reparameterizations.}
    \label{tab:full-opt-results}
    \begin{tabular}{r|*{9}{S[table-format=5,tight-spacing=true]}}
        \toprule
        Optimizer & {\orig} & {\most{10}} & {\most{20}} & {\most{33}} & {\most{43}} & {\least{10}} & {\least{20}} & {\least{33}} & {\least{43}} \\
        \midrule
        1           &  319 & 1871 &  680 &  486 &  380 & 19918 & 18414 & 14588 & 2650 \\
        2           &  308 & 1876 &  680 &  497 &  387 & 19918 & 18431 & 14588 & 2622 \\
        3           &  351 & 1869 &  682 &  419 &  392 & 19918 & 18431 & 14589 & 2703 \\
        4           &  334 & 2428 &  488 &  415 &  327 & 19918 & 18414 & 14588 & 2607 \\
        5           &  409 & 1873 &  490 &  417 &  343 & 19918 & 18413 & 14596 & 2650 \\
        6           &  343 & 1873 &  684 &  479 &  489 & 19918 & 18414 & 14588 & 2621 \\
        7           &  352 & 1879 &  681 &  491 &  355 & 19918 & 18415 & 14589 & 2637 \\
        8           &  364 & 1875 &  682 &  471 &  412 & 19918 & 18414 & 14588 & 2901 \\
        9           &  410 & 1869 &  684 &  478 &  336 & 19918 & 18413 & 14587 & 2747 \\
        10          &  334 & 1195 &  486 &  484 &  351 & 19918 & 18432 & 14642 & 2597 \\
        11          &  337 & 1874 &  679 &  424 &  390 & 19918 & 18430 & 14589 & 2616 \\
        12          &  314 & 1871 &  487 &  530 &  397 & 19918 & 18413 & 14589 & 2617 \\
        13          &  351 & 1875 &  681 &  378 &  389 & 19918 & 18415 & 14591 & 3097 \\
        14          &  316 & 1870 &  489 &  577 &  356 & 19918 & 18413 & 14589 & 2617 \\
        15          &  347 & 1888 &  681 &  376 &  330 & 19918 & 18412 & 14588 & 2606 \\
        16          &  342 & 1194 &  485 &  424 &  388 & 19918 & 18412 & 14588 & 2616 \\
        \midrule
        Min.\       &  308 & 1194 &  485 &  376 &  327 & 19918 & 18412 & 14587 & 2597 \\
        Max.\       &  410 & 2428 &  684 &  577 &  489 & 19918 & 18432 & 14642 & 3097 \\
        Std.\ Dev.\ &   33 &  339 &   98 &   62 &   48 &     0 &     8 &    17 &  162 \\
        \bottomrule
    \end{tabular}
\end{table}

\newpage
\section{Sensitivity results as a function of kernel parameters}

\begin{figure}
    \centering
    \includegraphics{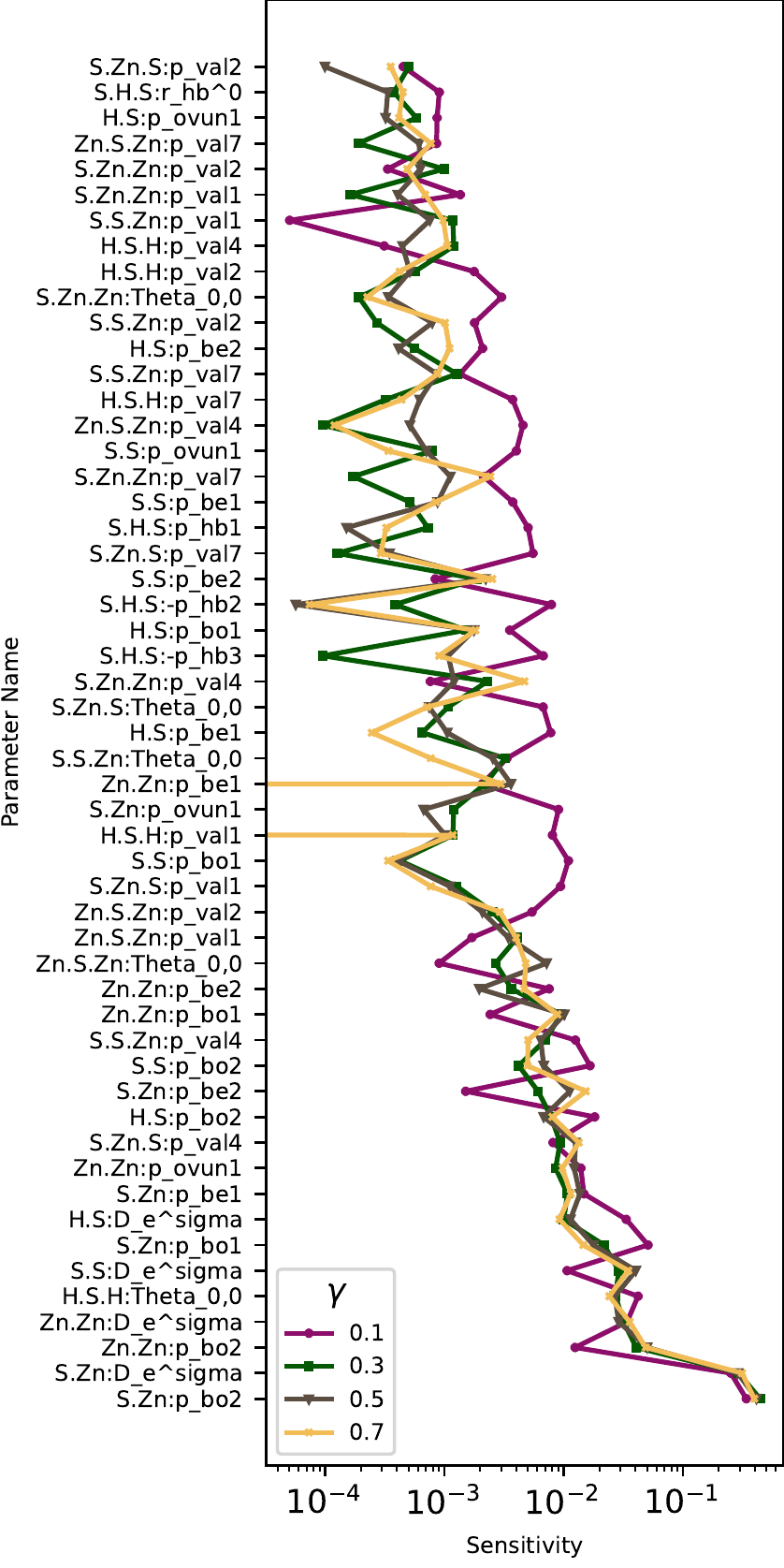}
    \caption{Average sensitivity values (over ten runs each) determined for each parameter using different $\gamma$ values for the CG kernel. Parameters are ordered by the average across kernels.}
    \label{fig:compare-kernel-param-values}
\end{figure}

Figure~\ref{fig:compare-kernel-param-values} shows sensitivity results using the CG kernel with different values for its shape parameter, $\gamma$.
The ReaxFF parameters are ordered by the average sensitivity value across all kernels so that none are given preference.
Each of the kernels is able to identify a similar ordering for the most sensitive parameters, however, very low and very high $\gamma$ values do not provide a wide-enough range of kernel values for the method to properly resolve out the sensitivities.
The results for $\gamma=0.1$ are very disperse across the ten repeats, with the orderings of the moderately sensitive parameters changing significantly between repeats.
This is because nearly all the kernel values are very close to zero, and thus hard to differentiate.
This leads to the average value shown in the figure being very different from the values calculated using larger $\gamma$ values.
The same issue is encountered for $\gamma=0.7$, but in this case the kernel values are very near to one.
One can see, for example, that the parameter \texttt{S.Zn:p\_ovun1}, which is moderately sensitive according to all the other kernels, has a sensitivity of zero.
This is primarily an artifact of the thresholding we use to truncate negative HSIC values to zero, but it occurs because the raw HSIC values are extremely small when using this large parameter value.
The moderate values of $\gamma=0.3$ and $\gamma=0.5$ produce similar results.
This illustrates that the ordering is robust, and that users need not overly trouble themselves with selecting kernel parameter values, as long as they are reasonable.

\newpage
\section{Calculation of adsorption energies}
\label{sec:calculation-of-adsorption-energies}

\begin{figure}[htpb]
    \centering
    \subcaptionbox{}{\includegraphics[width=0.49\linewidth]{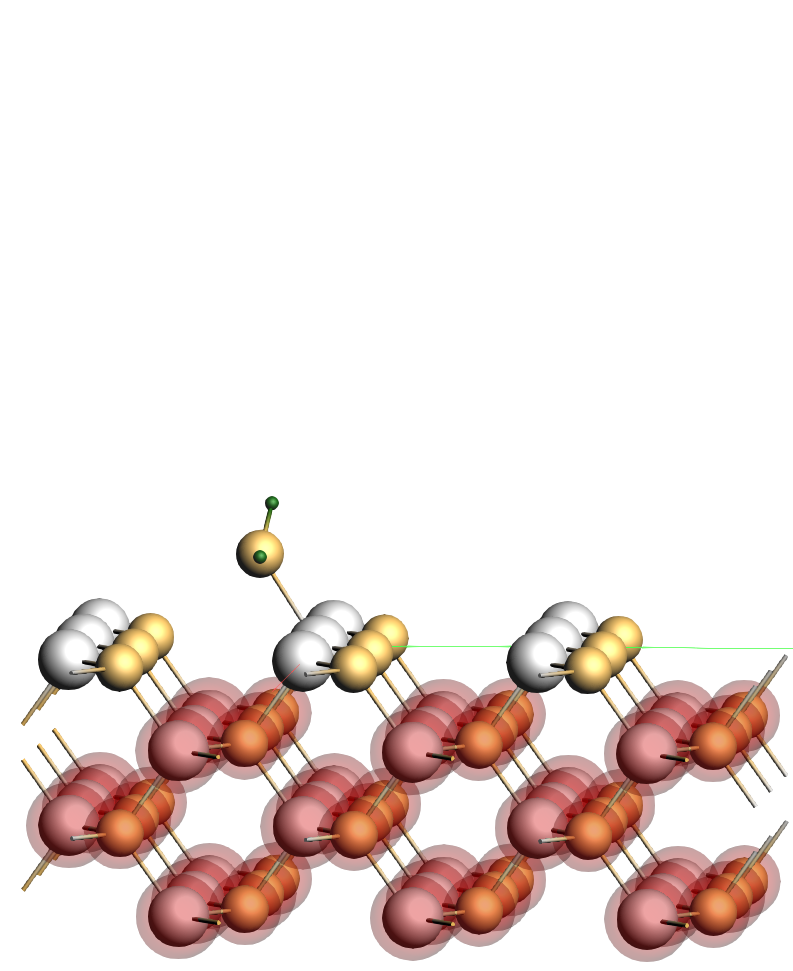}}
    \subcaptionbox{}{\includegraphics[width=0.49\linewidth]{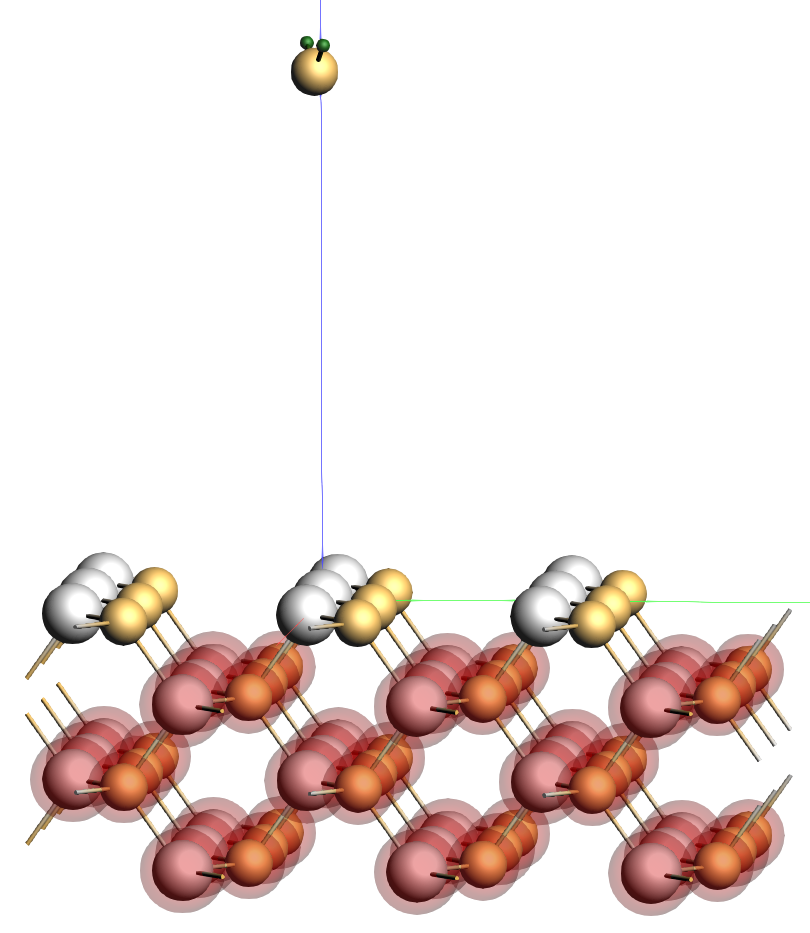}}
    \caption{Systems modelling an \hs molecule (a) on and (b) above a \zplane zincblende surface.}
    \label{fig:h2s_adsorption}
\end{figure}

Geometry optimizations were performed on the systems in Figure~\ref{fig:h2s_adsorption}.
In these optimizations the atoms haloed in red are fixed.
Adsorption energy was calculated via an in-cell method, i.e., the energy difference in energies of formation between the two systems:
\begin{equation}
    \label{eq:in-cell}
    \Delta E_\text{ads} = E_\text{slab+adsorbed molecule} - E_\text{slab+far away molecule}.
\end{equation}
The surfaces were constructed using pre-optimized lattice parameters.

\newpage
\section{Calculation of surface energies}
\label{sec:calculation-of-surface-energies}

\begin{figure}[htpb]
    \centering
    \subcaptionbox{Wurtzite \wplane}{\includegraphics[width=0.49\textwidth]{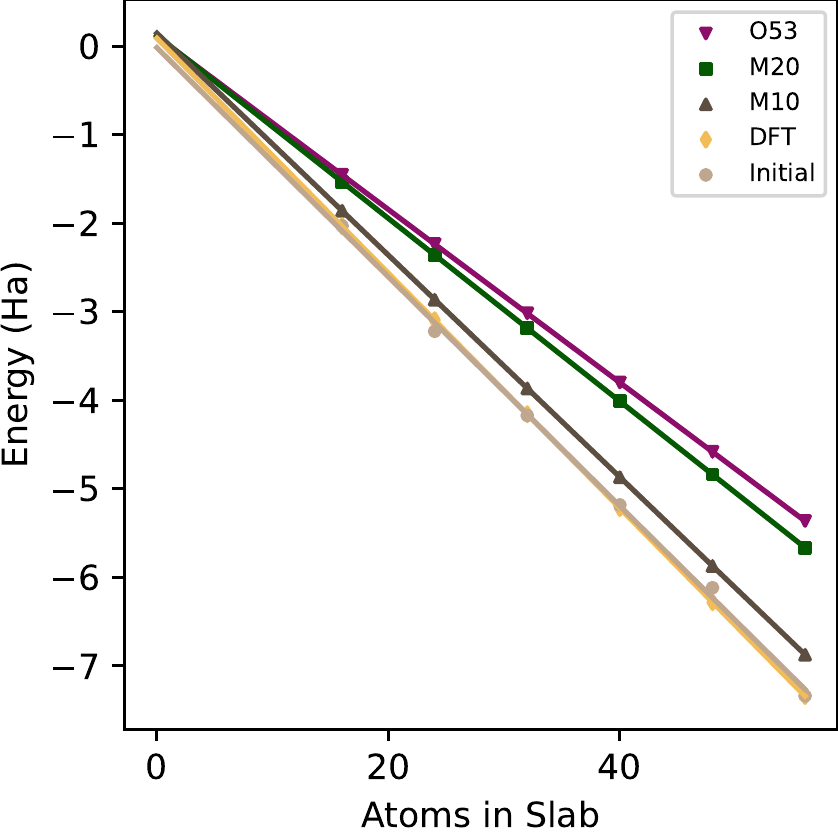}}
    \subcaptionbox{Zincblende \zplane}{\includegraphics[width=0.49\textwidth]{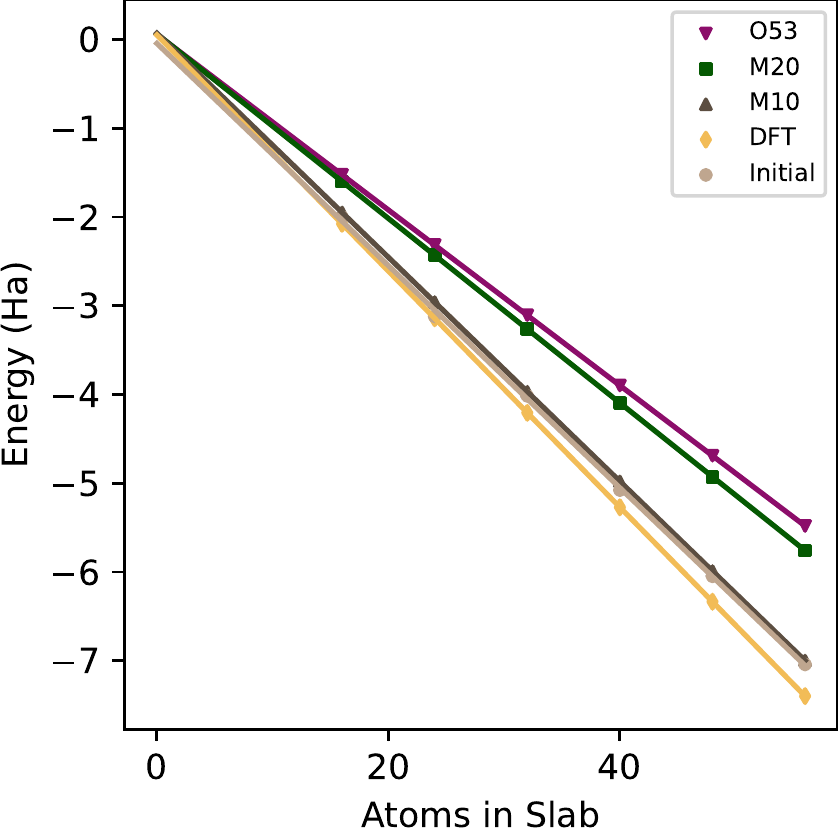}}
    \caption{Calculation of surface energies via linear regression.}
    \label{fig:linreg}
\end{figure}

Geometry optimizations were performed on slabs of varying thicknesses for \zplane zincblende and \wplane wurtzite.
The slabs were constructed from pre-optimized lattice parameters.
Energies were linearly regressed as a function of the number of atoms in the slab (see Figure~\ref{fig:linreg}).
Surface energies were calculated from the intercepts of these regressions using:
\begin{equation}
    \label{eq:surf_eng}
    E_\textrm{surf} = \frac{c}{2A},
\end{equation}
where $c$ is the intercept, and $A$ is the slab surface area, calculated as \SIlist{20.7;40.9}{\angstrom\squared} for \zplane zincblende and \wplane wurtzite respectively.
Table~\ref{tab:linreg} lists the calculated fits.

\begin{table*}[htbp]
    \centering
    \caption{Linear regressions ($E = mn + c$) of energies~($E$) of slabs of different thicknesses for the determination of surface energies for \zplane zincblende and \wplane wurtzite crystals, where $n$ is the number of atoms in the slab.}
    \label{tab:linreg}
        \begin{tabular}{lSSSS}
            \toprule
             & \multicolumn{2}{c}{Wurtzite} & \multicolumn{2}{c}{Zincblende} \\\cmidrule(r{2pt}){2-3}\cmidrule(l{2pt}){4-5}
             & $m$ & $c$ & $m$ & $c$ \\
            \midrule
            DFT       & -0.133 & 0.089  & -0.133 & 0.051 \\
            Initial   & -0.129 & -0.017 & -0.125 & -0.044\\
            \orig     & -0.098 & 0.110  & -0.099 & 0.059 \\
            \most{20} & -0.103 & 0.120  & -0.104 & 0.063 \\
            \most{10} & -0.125 & 0.146  & -0.126 & 0.071 \\
            \bottomrule
        \end{tabular}
\end{table*}

\newpage
\section{Full surface and adsorption energy results}
\label{sec:full-surface-and-adsorption-energy-results}

\begin{table*}[htpb]
    \centering
    \caption{SSE, adsorption energy and surface energy results for the best force fields found by each of the sixteen \orig and \most{20} optimizers. Row/optimizer with lowest loss values are in bold.}
    \label{tab:full-validation}
    \begin{threeparttable}
        \begin{tabular}{*{9}r}
            \toprule
             & \multicolumn{4}{c}{\orig} & \multicolumn{4}{c}{\most{20}} \\
            \cmidrule(r{2pt}){2-5}\cmidrule(l{2pt}){6-9}
             & & & \multicolumn{2}{c}{$E_\mathrm{surf}$ (\si{\joule\per\meter\squared})} & & & \multicolumn{2}{c}{$E_\mathrm{surf}$ (\si{\joule\per\meter\squared})} \\
            \cmidrule{4-5}\cmidrule{8-9}
            Opt.\tnote{*}\ & SSE\tnote{\dag} & $E_\mathrm{ads}$ (eV) & W. \wplane\tnote{\ddag} & Z. \zplane\tnote{\textparagraph} & SSE\tnote{\dag} & $E_\mathrm{ads}$ (eV) & W. \wplane\tnote{\ddag} & Z. \zplane\tnote{\textparagraph} \\
            \midrule
            1  & 319 & -0.762 & 0.619 & 0.659 & 680 & -0.839 & 0.597 & 0.576 \\
            2 & \textbf{308} & \textbf{-0.952} & \textbf{0.585} & \textbf{0.617} & 680 & 0.593 & 0.571 & -0.828 \\
            3  & 351 & -0.783 & 0.665 & 0.650 & 682 & -0.820 & 0.561 & 0.544 \\
            4  & 334 & -0.764 & 0.656 & 0.691 & 488 & -0.886 & 0.662 & 0.683 \\
            5  & 409 & -0.426 & 0.305 & 0.358 & 490 & -0.848 & 0.654 & 0.674 \\
            6  & 343 & -0.806 & 0.620 & 0.650 & 684 & -0.812 & 0.556 & 0.538 \\
            7  & 352 & -0.790 & 0.657 & 0.636 & 681 & -0.832 & 0.596 & 0.575 \\
            8  & 364 & -0.762 & 0.645 & 0.622 & 682 & -0.824 & 0.591 & 0.568 \\
            9  & 410 & -0.748 & 0.660 & 0.734 & 684 & -0.814 & 0.582 & 0.560 \\
            10 & 334 & -1.136 & 0.582 & 0.629 & 486 & -0.861 & 0.648 & 0.670 \\
            11 & 337 & -0.780 & 0.575 & 0.598 & 679 & -0.828 & 0.592 & 0.569 \\
            12 & 314 & -0.727 & 0.613 & 0.643 & 487 & -0.861 & 0.654 & 0.673 \\
            13 & 351 & -0.720 & 0.574 & 0.599 & 681 & -0.827 & 0.595 & 0.572 \\
            14 & 316 & -0.766 & 0.678 & 0.703 & 489 & -0.863 & 0.656 & 0.676 \\
            15 & 347 & -0.775 & 0.679 & 0.709 & 681 & -0.827 & 0.583 & 0.561 \\
            16 & 342 & 0.556 & 0.598 & -0.767 & \textbf{485} & \textbf{-0.852} & \textbf{0.641} & \textbf{0.661} \\
            \midrule
            Ave.\ & 346 & 0.604 & 0.631 & -0.779 & 609 & 0.610 & 0.605 & -0.839 \\
            Min.\ & 308 & 0.305 & 0.358 & -1.136 & 485 & 0.556 & 0.538 & -0.886 \\
            Max.\ & 410 & 0.679 & 0.734 & -0.426 & 684 & 0.662 & 0.683 & -0.812 \\
            Std.\ & 29 & 0.090 & 0.084 & 0.139 & 97 & 0.036 & 0.056 & 0.021 \\
            \bottomrule
        \end{tabular}
        \begin{tablenotes}
            \item[*] Optimizer number
            \item[\dag] Sum of square errors (see Equation~7)
            \item[\ddag] Wurtzite \wplane surface
            \item[\textparagraph] Zincblende \zplane surface
        \end{tablenotes}
    \end{threeparttable}
\end{table*}

\newpage
\section{Composition of the training set}
\label{sec:composition-of-the-training-set}

This appendix contains an description of all the (alphabetically sorted) jobs from which the training data is extracted.
For each job we provide:
\begin{itemize}
    \item a unique job ID which will be used to reference items;
    \item image of the system starting geometry;
    \item description of the job run;
    \item details of the training data extracted from the job.
\end{itemize}

DFT calculations for the reference data were run using BAND with the PBEsol functional together with DZ or DZP basis sets.
For the volume scans, the k-space quality was set to `Good', corresponding to circa \num{27} k-points per \si{\angstrom\cubed}.

\begin{minipage}{\linewidth}
    \bigskip
    \begin{description}
        \item[Job ID: \texttt{anglescan\_h2s\_pbesol}] \leavevmode\\
            \includegraphics[width=0.3\linewidth]{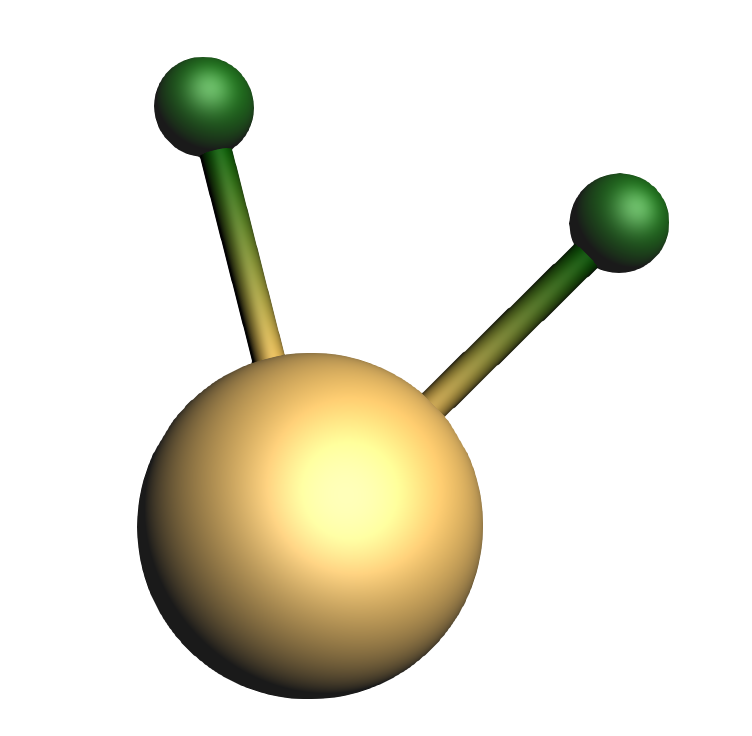}
            \begin{description}
                \item[Job description] \leavevmode
                    \begin{itemize}
                        \item \num{10} point angle scan from \SIrange{60}{180}{\degree}.
                        \item Max \num{30} point geometry optimization at each angle
                    \end{itemize}
                \item[Extracted training data] \leavevmode\\
                    \begin{itemize}
                        \item Energies for each of the \num{10} geometries
                    \end{itemize}
            \end{description}
    \end{description}
\end{minipage}
\begin{minipage}{\linewidth}
    \bigskip
    \begin{description}
        \item[Job ID: \texttt{band\_110}] \leavevmode\\
            \includegraphics[width=0.3\linewidth]{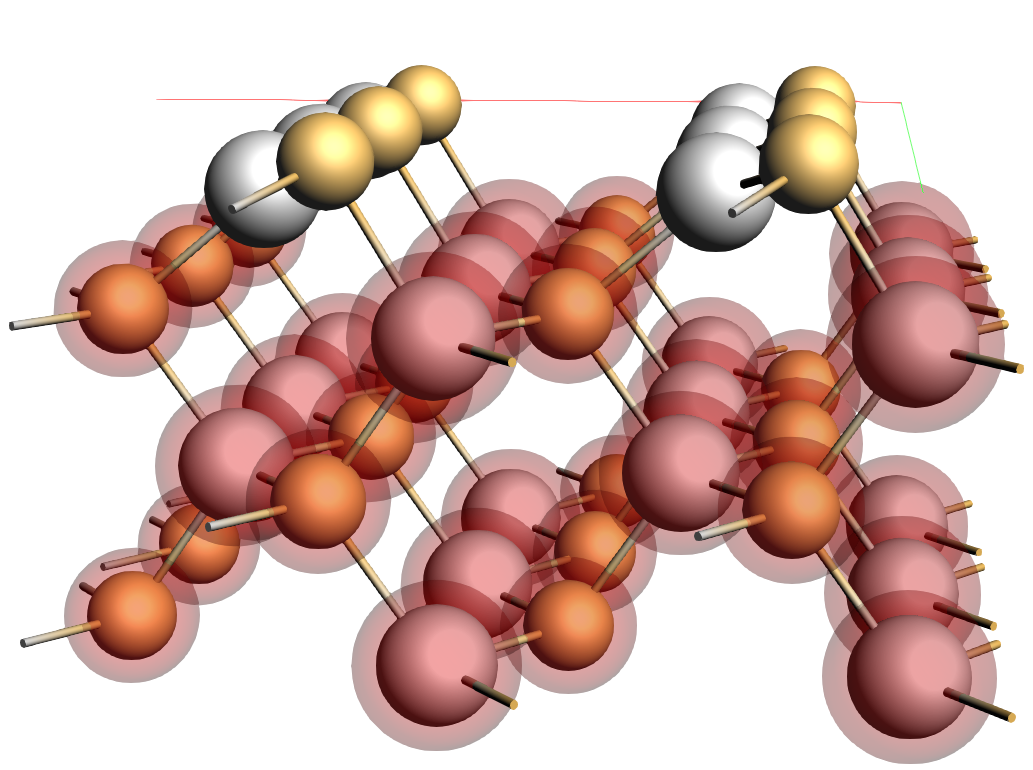}
            \begin{description}
                \item[Job description] \leavevmode
                    \begin{itemize}
                        \item Max \num{30} point geometry optimization
                        \item Atoms haloed in red are fixed
                    \end{itemize}
                \item[Extracted training data] \leavevmode
                    \begin{itemize}
                        \item Energy difference between \texttt{band\_110} and \texttt{band\_distorted\_clean\_110}
                        \item \hs adsorption energy: (\texttt{band\_h2s\_110} - \texttt{band\_110} - \texttt{band\_h2s})
                    \end{itemize}
            \end{description}
    \end{description}
\end{minipage}
\begin{minipage}{\linewidth}
    \bigskip
    \begin{description}
        \item[Job ID: \texttt{band\_110\_noconstraints}] \leavevmode\\
            \includegraphics[width=0.3\linewidth]{jobs_band_110}
            \begin{description}
                \item[Job description] \leavevmode
                    \begin{itemize}
                        \item Max \num{30} point geometry optimization
                        \item No atom positions are fixed
                    \end{itemize}
                \item[Extracted training data] \leavevmode
                    \begin{itemize}
                        \item Atomic charges (\num{48} total)
                        \item \num{5} angles from the interior and surface of the slab
                        \item \num{1} S-Zn-S-Zn dihedral angle
                        \item \num{9} interatomic distances from the interior and surface of the slab
                        \item Energy difference between \texttt{band\_110\_noconstraints} and \texttt{wurtzite\_sp}
                    \end{itemize}
            \end{description}
    \end{description}
\end{minipage}
\begin{minipage}{\linewidth}
    \bigskip
    \begin{description}
        \item[Job ID: \texttt{band\_distorted\_ads\_110}] \leavevmode\\
            \includegraphics[width=0.3\linewidth]{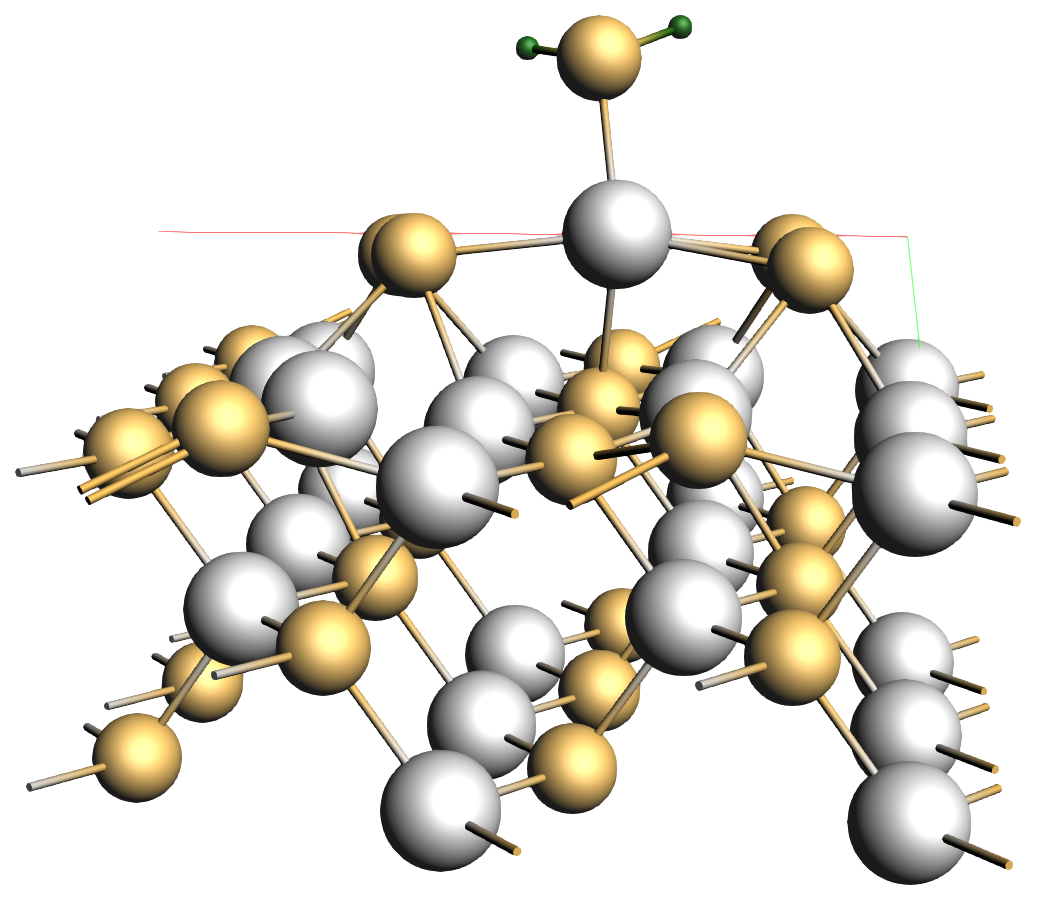}
            \begin{description}
                \item[Job description] \leavevmode
                    \begin{itemize}
                        \item Single point calculation
                        \item Geometry obtained from DFT geometry optimization of MD snapshot
                    \end{itemize}
                \item[Extracted training data] \leavevmode
                    \begin{itemize}
                        \item Atomic forces (\num{153} total)
                        \item Energy difference between \texttt{band\_distorted\_ads\_110} and \texttt{band\_h2s\_110}
                    \end{itemize}
            \end{description}
    \end{description}
\end{minipage}
\begin{minipage}{\linewidth}
    \bigskip
    \begin{description}
        \item[Job ID: \texttt{band\_distorted\_clean\_110}] \leavevmode\\
            \includegraphics[width=0.3\linewidth]{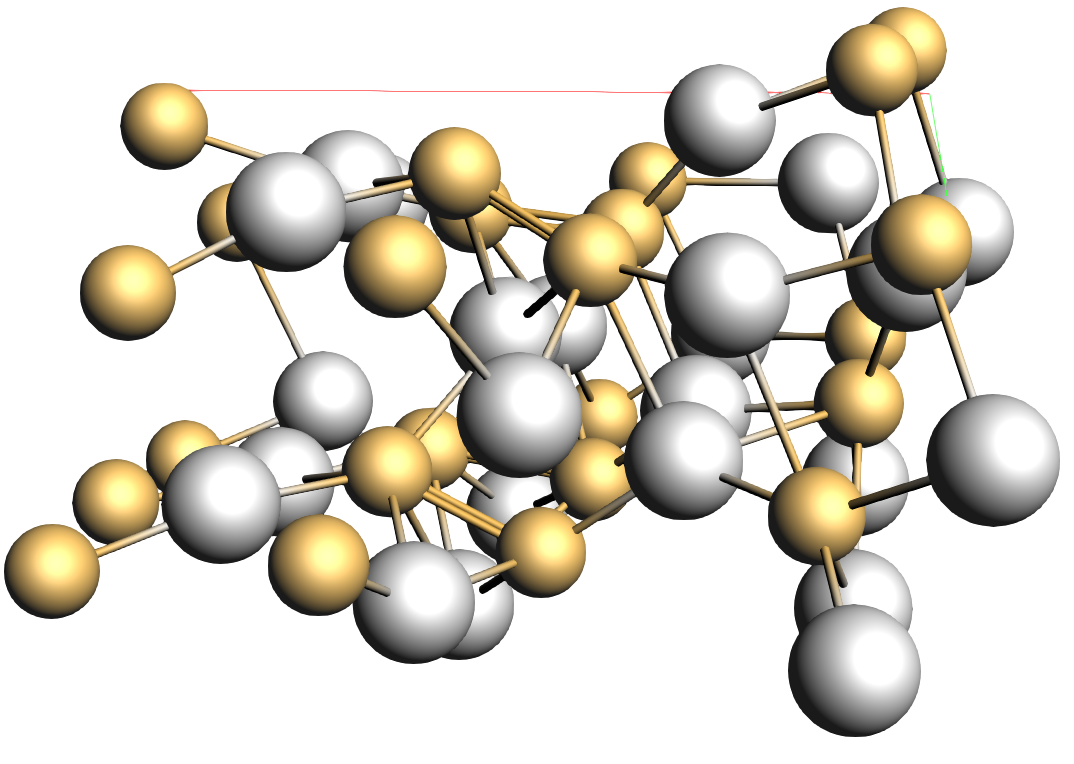}
            \begin{description}
                \item[Job description] \leavevmode
                    \begin{itemize}
                        \item Single point calculation
                        \item Geometry obtained from DFT geometry optimization of MD snapshot
                    \end{itemize}
                \item[Extracted training data] \leavevmode
                    \begin{itemize}
                        \item Atomic forces (\num{144} total)
                        \item Energy difference between \texttt{band\_distorted\_clean\_110} and \texttt{band\_110}
                    \end{itemize}
            \end{description}
    \end{description}
\end{minipage}
\begin{minipage}{\linewidth}
    \bigskip
    \begin{description}
        \item[Job ID: \texttt{band\_h2s}] \leavevmode\\
            \newline
            \includegraphics[width=0.3\linewidth]{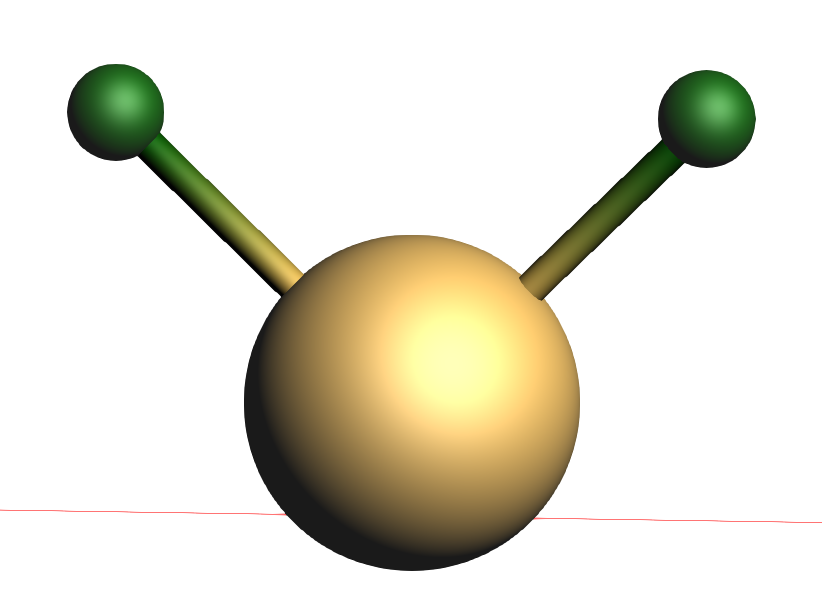}
            \begin{description}
                \item[Job description] \leavevmode
                    \begin{itemize}
                        \item Max \num{30} point geometry optimization
                    \end{itemize}
                \item[Extracted training data] \leavevmode
                    \begin{itemize}
                        \item Atomic charges (\num{3} total)
                        \item Bond angle
                        \item H-S bond distance
                        \item \hs adsorption energy: (\texttt{band\_h2s\_110} - \texttt{band\_110} - \texttt{band\_h2s})
                    \end{itemize}
            \end{description}
    \end{description}
\end{minipage}
\begin{minipage}{\linewidth}
    \bigskip
    \begin{description}
        \item[Job ID: \texttt{band\_h2s\_110}] \leavevmode\\
            \newline
            \includegraphics[width=0.3\linewidth]{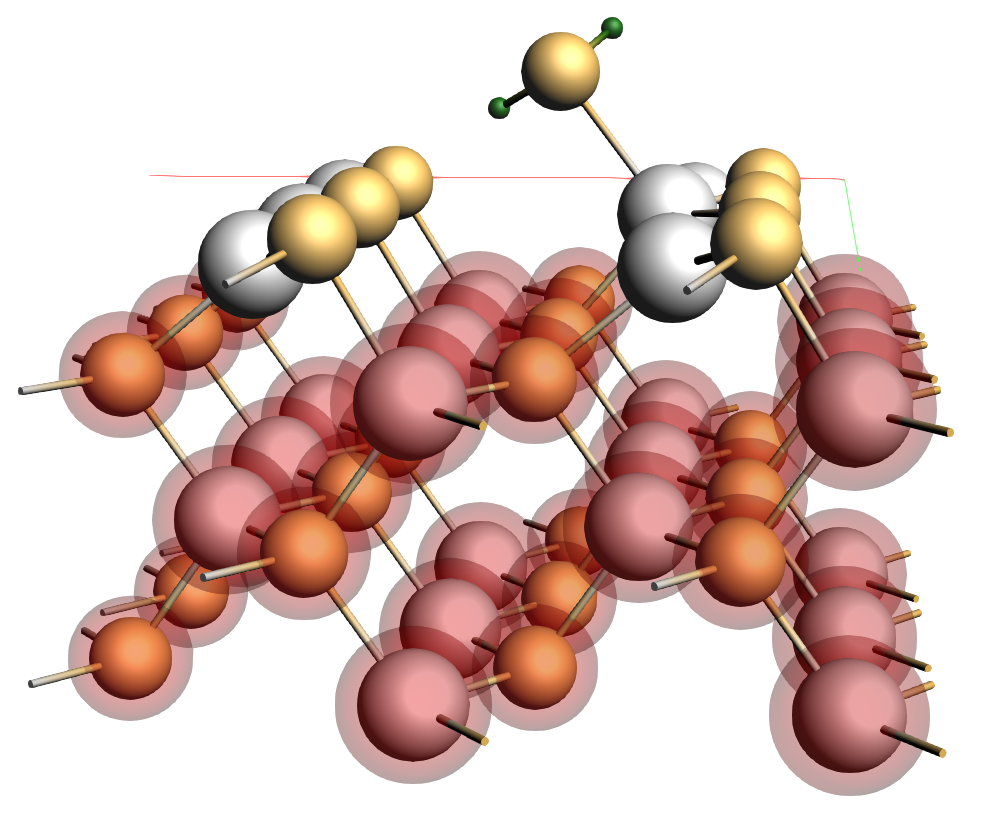}
            \begin{description}
                \item[Job description] \leavevmode
                    \begin{itemize}
                        \item Max \num{30} point geometry optimization
                        \item Atoms haloed in red are fixed
                    \end{itemize}
                \item[Extracted training data] \leavevmode
                    \begin{itemize}
                        \item Atomic charges (\num{51} total)
                        \item S-Zn adsorption distance
                        \item H-S adsorption distance
                        \item \hs adsorption energy: (\texttt{band\_h2s\_110} - \texttt{band\_110} - \texttt{band\_h2s})
                        \item Energy difference between \texttt{band\_distorted\_ads\_110} and \texttt{band\_h2s\_110}
                    \end{itemize}
            \end{description}
    \end{description}
\end{minipage}
\begin{minipage}{\linewidth}
    \bigskip
    \begin{description}
        \item[Job ID: \texttt{bondscan\_h2s\_pbesol}] \leavevmode\\
            \newline
            \includegraphics[width=0.3\linewidth]{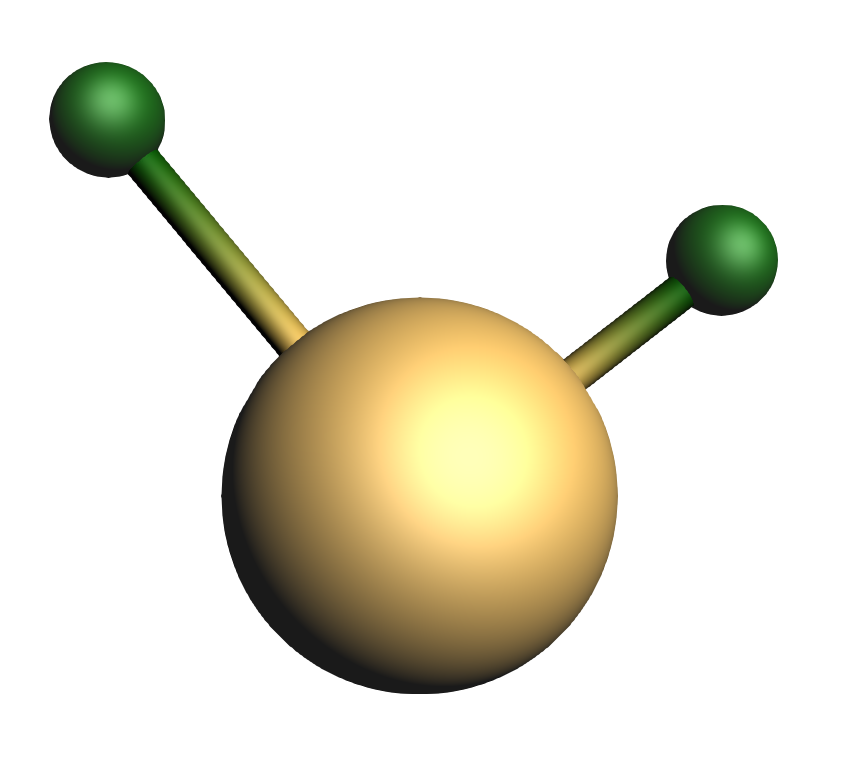}
            \begin{description}
                \item[Job description] \leavevmode
                    \begin{itemize}
                        \item \num{10} point bond length scan from \SIrange{1.1}{1.6}{\angstrom}
                        \item Max \num{30} point geometry optimization at each bond length
                    \end{itemize}
                \item[Extracted training data] \leavevmode
                    \begin{itemize}
                        \item Energies for each of the \num{10} geometries
                    \end{itemize}
            \end{description}
    \end{description}
\end{minipage}
\begin{minipage}{\linewidth}
    \bigskip
    \begin{description}
        \item[Job ID: \texttt{rocksalt}] \leavevmode\\
            \newline
            \includegraphics[width=0.3\linewidth]{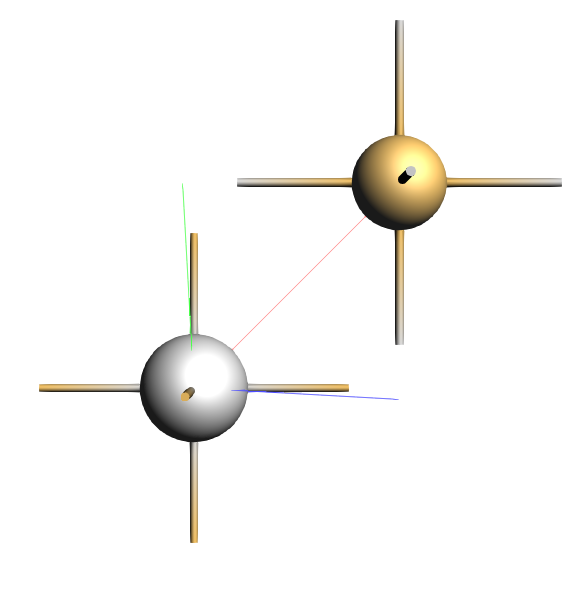}
            \begin{description}
                \item[Job description] \leavevmode
                    \begin{itemize}
                        \item \num{5} point cell volume scan, scaled from \SIrange{85}{125}{\percent}
                        \item Single point energy calculation for each geometry
                    \end{itemize}
                \item[Extracted training data] \leavevmode
                    \begin{itemize}
                        \item Energies for each of the \num{5} geometries
                    \end{itemize}
            \end{description}
    \end{description}
\end{minipage}
\begin{minipage}{\linewidth}
    \bigskip
    \begin{description}
        \item[Job ID: \texttt{rocksalt\_sp}] \leavevmode\\
            \newline
            \includegraphics[width=0.3\linewidth]{jobs_rocksalt}
            \begin{description}
                \item[Job description] \leavevmode
                    \begin{itemize}
                        \item Single point calculation
                    \end{itemize}
                \item[Extracted training data] \leavevmode
                    \begin{itemize}
                        \item Atomic charges (\num{2} total)
                        \item Energy difference (\texttt{rocksalt\_sp} - 0.03125\texttt{sulfur} - 0.5\texttt{Zn})
                        \item Energy difference between \texttt{rocksalt\_sp} and \texttt{zincblende\_sp}
                    \end{itemize}
            \end{description}
    \end{description}
\end{minipage}
\begin{minipage}{\linewidth}
    \bigskip
    \begin{description}
        \item[Job ID: \texttt{sulfur}] \leavevmode\\
            \newline
            \includegraphics[width=0.3\linewidth]{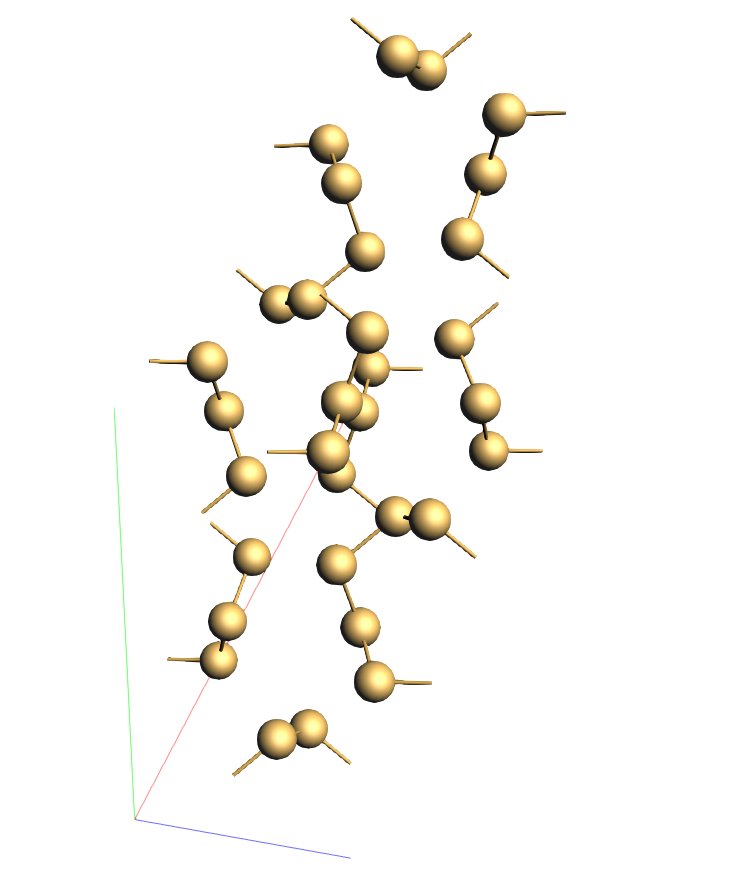}
            \begin{description}
                \item[Job description] \leavevmode
                    \begin{itemize}
                        \item Single point calculation
                    \end{itemize}
                \item[Extracted training data] \leavevmode
                    \begin{itemize}
                        \item Energy difference (\texttt{rocksalt\_sp} - 0.03125\texttt{sulfur} - 0.5\texttt{Zn})
                        \item Energy difference (\texttt{zincblende\_sp} - 0.03125\texttt{sulfur} - 0.5\texttt{Zn})
                        \item Energy difference (0.5\texttt{wurtzite\_sp} - 0.03125\texttt{sulfur} - 0.5\texttt{Zn})
                    \end{itemize}
            \end{description}
    \end{description}
\end{minipage}
\begin{minipage}{\linewidth}
    \bigskip
    \begin{description}
        \item[Job ID: \texttt{wurtzite}] \leavevmode\\
            \newline
            \includegraphics[width=0.3\linewidth]{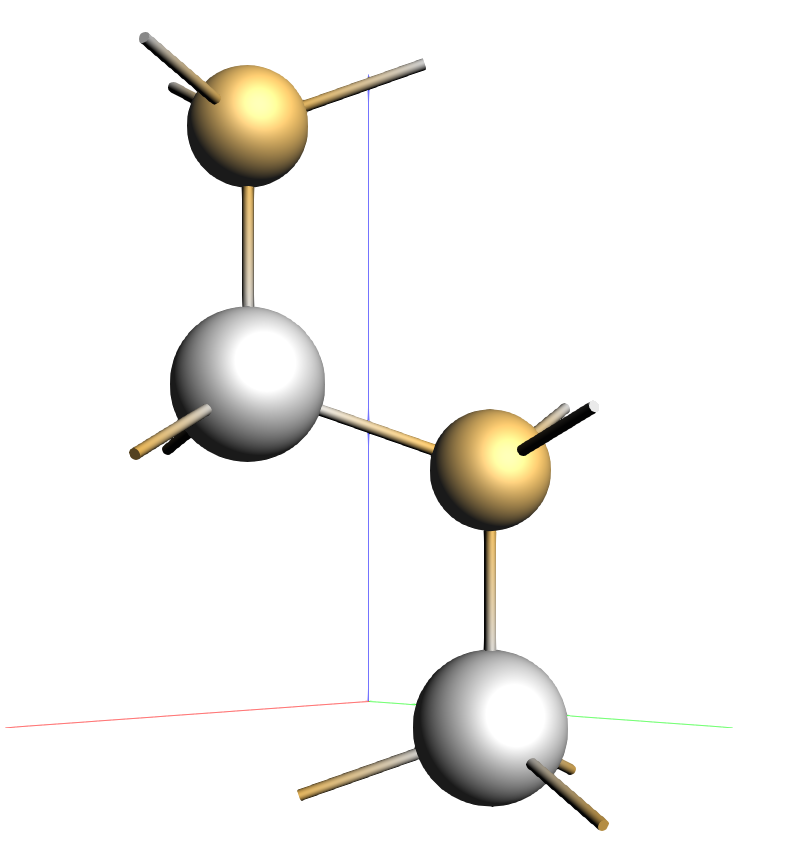}
            \begin{description}
                \item[Job description] \leavevmode
                    \begin{itemize}
                        \item \num{5} point cell volume scan, scaled from \SIrange{85}{125}{\percent}
                        \item Single point energy calculation for each geometry
                    \end{itemize}
                \item[Extracted training data] \leavevmode
                    \begin{itemize}
                        \item Energies for each of the \num{5} geometries
                    \end{itemize}
            \end{description}
    \end{description}
\end{minipage}
\begin{minipage}{\linewidth}
    \bigskip
    \begin{description}
        \item[Job ID: \texttt{wurtzite\_sp}] \leavevmode\\
            \newline
            \includegraphics[width=0.3\linewidth]{jobs_wurtzite}
            \begin{description}
                \item[Job description] \leavevmode
                    \begin{itemize}
                        \item Single point calculation
                    \end{itemize}
                \item[Extracted training data] \leavevmode
                    \begin{itemize}
                        \item Atomic charges (\num{4} total)
                        \item Energy difference between \texttt{band\_110\_noconstraints} and \texttt{wurtzite\_sp}
                        \item Energy difference between \texttt{zincblende\_sp} and \texttt{wurtzite\_sp}
                        \item Energy difference (0.5\texttt{wurtzite\_sp} - 0.03125\texttt{sulfur} - 0.5\texttt{Zn})
                    \end{itemize}
            \end{description}
    \end{description}
\end{minipage}
\begin{minipage}{\linewidth}
    \bigskip
    \begin{description}
        \item[Job ID: \texttt{zincblende}] \leavevmode\\
            \newline
            \includegraphics[width=0.3\linewidth]{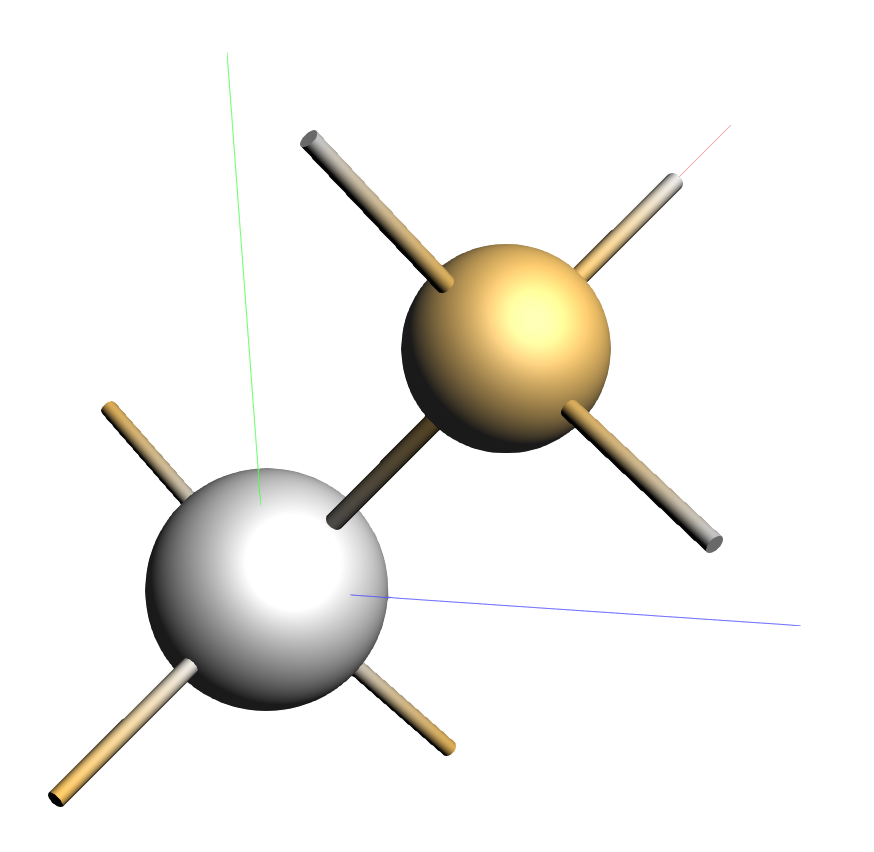}
            \begin{description}
                \item[Job description] \leavevmode
                    \begin{itemize}
                        \item \num{5} point cell volume scan, scaled from \SIrange{85}{125}{\percent}
                        \item Single point energy calculation for each geometry
                    \end{itemize}
                \item[Extracted training data] \leavevmode
                    \begin{itemize}
                        \item Energies for each of the \num{5} geometries
                    \end{itemize}
            \end{description}
    \end{description}
\end{minipage}
\begin{minipage}{\linewidth}
    \bigskip
    \begin{description}
        \item[Job ID: \texttt{zincblende\_sp}] \leavevmode\\
            \newline
            \includegraphics[width=0.3\linewidth]{jobs_zincblende}
            \begin{description}
                \item[Job description] \leavevmode
                    \begin{itemize}
                        \item Single point calculation
                    \end{itemize}
                \item[Extracted training data] \leavevmode
                    \begin{itemize}
                        \item Atomic charges (\num{2} total)
                        \item Energy difference between \texttt{rocksalt\_sp} and \texttt{zincblende\_sp}
                        \item Energy difference between \texttt{zincblende\_sp} and \texttt{wurtzite\_sp}
                        \item Energy difference (\texttt{zincblende\_sp} - 0.03125\texttt{sulfur} - 0.5\texttt{Zn})
                    \end{itemize}
            \end{description}
    \end{description}
\end{minipage}
\begin{minipage}{\linewidth}
    \bigskip
    \begin{description}
        \item[Job ID: \texttt{Zn}] \leavevmode\\
            \newline
            \includegraphics[width=0.3\linewidth]{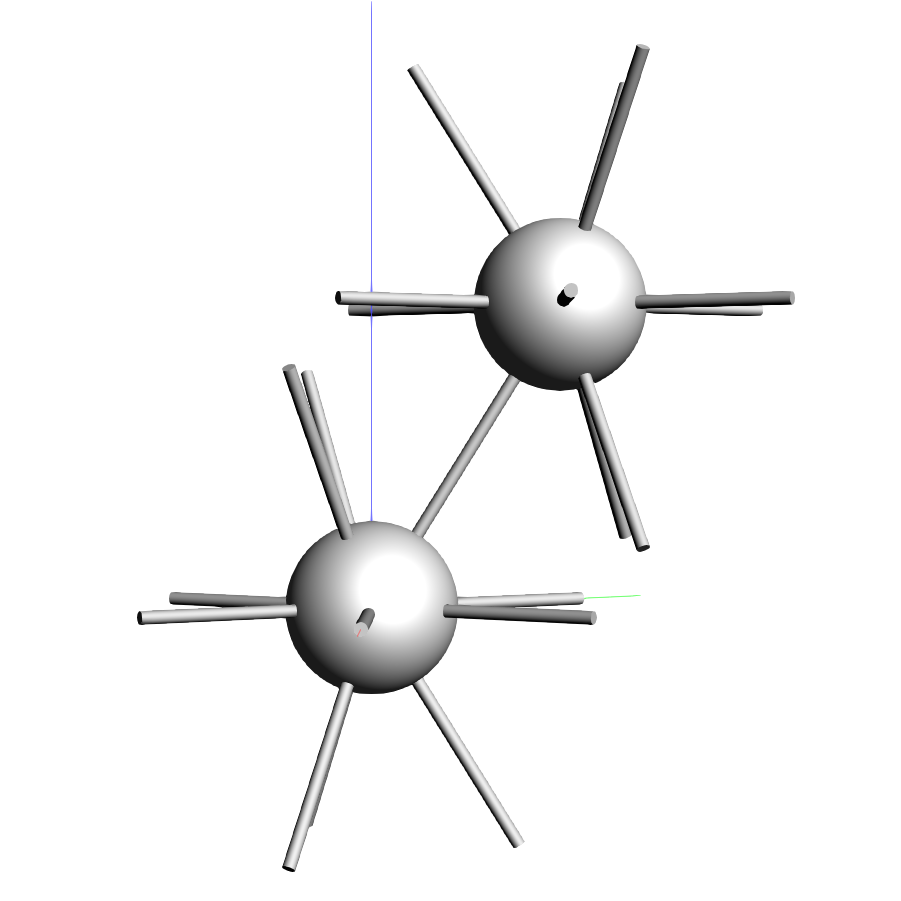}
            \begin{description}
                \item[Job description] \leavevmode
                    \begin{itemize}
                        \item Single point calculation
                    \end{itemize}
                \item[Extracted training data] \leavevmode
                    \begin{itemize}
                        \item Atomic charges (\num{2} total)
                        \item Energy difference (0.5\texttt{wurtzite\_sp} - 0.03125\texttt{sulfur} - 0.5\texttt{Zn})
                        \item Energy difference (\texttt{zincblende\_sp} - 0.03125\texttt{sulfur} - 0.5\texttt{Zn})
                        \item Energy difference (\texttt{rocksalt\_sp} - 0.03125\texttt{sulfur} - 0.5\texttt{Zn})
                    \end{itemize}
            \end{description}
    \end{description}
\end{minipage}